\documentclass[%
english,
reprint,
 aps,pre
]{revtex4-1}

\usepackage[latin9]{inputenc}
\usepackage{units}
\usepackage{amsmath}
\usepackage{amssymb}
\usepackage{graphicx}
\usepackage{bm}
\usepackage{esint}
\usepackage{babel}
\usepackage{color}
\usepackage{xcolor}
\usepackage{tikz}

\definecolor{color4}{rgb}{0.701517, 0.172918, 0.174348}
\definecolor{color10}{rgb}{0.734303, 0.508574, 0.693384}

\definecolor{mygreen}{rgb}{0.0, 0.5, 0.0}

\begin{document}

%\preprint{APS/123-QED}

\title{Foldable Cones as a Framework for Nonrigid Origami}
\author{I. Andrade-Silva}
 \email{ignacio.andrade@ens-lyon.fr}
\affiliation{Universit\'e de Lyon, Ecole Normale Sup\'erieure de Lyon, Universit\'e Claude Bernard, CNRS, Laboratoire de Physique, F-69342 Lyon, France}
\author{M. Adda-Bedia}
\affiliation{Universit\'e de Lyon, Ecole Normale Sup\'erieure de Lyon, Universit\'e Claude Bernard, CNRS, Laboratoire de Physique, F-69342 Lyon, France}
\author{M. A. Dias}
\affiliation{Department of Engineering, Aarhus University, 8000 Aarhus C, Denmark\\
Aarhus University Centre for Integrated Materials Research--$\mathrm{iMAT}$, 8000 Aarhus C, Denmark}
\date{\today}

\begin{abstract}
The study of origami-based mechanical metamaterials usually focuses on the kinematics of deployable structures made of an assembly of rigid flat plates connected by hinges. When the elastic response of each panel is taken into account, novel behaviors take place, as in the case of foldable cones ($f$-cones): circular sheets decorated by radial creases around which they can fold. These structures exhibit bistability, in the sense that they can snap-through from one metastable configuration to another. In this work, we study the elastic behavior of isometric $f$-cones for any deflection and crease mechanics, which introduce nonlinear corrections to a linear model studied previously. Furthermore, we test the inextensibility hypothesis by means of a continuous numerical model that includes both the extended nature of the creases, stretching and bending deformations of the panels. The results show that this phase field-like model could become an efficient numerical tool for the study of realistic origami structures.

%\begin{description}
%\end{description}
\end{abstract}

\pacs{Valid PACS appear here}

\maketitle

The basic premise of origami, the ancient Japanese art of paper folding, is to obtain a complex 3-dimensional structure starting from a 2-dimensional sheet to which a network of creases is imprinted. 
Despite the simplicity of this idea, in recent years, the field of mechanical metamaterials has sought inspiration from origami~\cite{Schenk3276,wei2013geometric} in the search of smart-materials with a vast range of functionality such as deployability of large membranes~\cite{miura1985method}, shape changing structures~\cite{dias2012geometric,dudte2016programming}, and tunable mechanical and thermal properties~\cite{silverberg2014using,waitukaitis2015origami,boatti2017origami,PhysRevLett.122.155501}, just to name a few. 
In practice, many of these applications are constrained to situations which origami structures are made from assemblies of flat rigid plates connected by hingelike creases.
In such situations, the geometrically accessible configurations are fully determined by the crease network, while the structural response is a result of the crease network and the crease mechanics~\cite{hanna2014waterbomb,brunck2016elastic,chen2015origami}.
By contrast, when the elastic response of the plates (mainly bending) is taken into account, a variety of new behaviors may emerge.
In this case, the elastic response of the structure is determined by the competition between the flexural stiffness of the panels $B$ and the torsional rigidity of the creases $k$.
The length $L^* \equiv B/k$, called origami length, determines wether the deformation of a non-rigid origami is bending or crease dominated~\cite{lechenault2014mechanical}.
If $l$ is the typical size of the facets, when $l\ll L^*$, the deformation is governed by the change on the folding angles, while if $l\gg L^*$, the deformation is governed by the bending of the panels.

However, suitable analytical models capturing the elastic regime of non-rigid origami still remains for the most part unexplored.
Foldable cones~\cite{lechenault2015generic}, or $f$-cones, are the simplest single-vertex non-rigid origami in which the elasticity of the plates is relevant.
$f$-cones are elastic sheets decorated by straight creases meeting at a single vertex around which they are folded.
As a first approximation, these sheets are assumed to be inextensible.
This results in a family of various umbrella-like motifs whose equilibrium shapes depend on the crease pattern imprinted in the flat configuration of the sheet and the mechanical response of the creases.
Regardless of the initial crease pattern, these structures exhibit bistability in the sense that they can mechanically snap through from one metastable configuration to another of higher elastic energy.

The $f$-cone belongs to a larger family of singularities emerging on sheets subjected to isometrical deformations~\cite{huffman1976curvature,seffen2016fundamental,guven2011conical}. 
In many situations, the elastic energy in a thin elastic sheet can localize in a single point leading to conical dislocation.
The most fundamental example is the so-called $d$-cone~\cite{ben1997crumpled,cerda1998conical}, a conical singularity observed when crumpling an elastic thin sheet. 
The bending energy of the defect diverges logarithmically as one gets closer to the vertex.
In a more realistic situation, these divergencies are regularized if the inextensibility constraint is relaxed, thus leading to stretching and plastic deformations close to the vertex~\cite{cerda1999conical}. 

The bistable behavior of $f$-cones was investigated in~\cite{lechenault2015generic} and a model was proposed to describe their equilibrium shapes in the limit of small deflection and infinitely stiff creases.
We will refer their model as the linear model for $f$-cones, as it relies on the approximation of small deflections which allows to write the curvature of the surface as a linear function of the vertical component of the displacement.
This system has motivated the study of other similar problems such as the bistable behavior of creased strips~\cite{walker2018shape}. In this last work, the authors proposed a discrete model based on the Gauss map of several creases meeting at the vertex. 
In the limit of infinite creases, the linear version of an $f$-cone with two creases is recovered.
The discrete model that is based on the Gauss map has limitations, as it only predicts the final shape of real sheets well for small deflections, while important discrepancies with experiments are observed for large deflections. 
Although, these discrepancies may be attributed to the existence of stretching in real sheets, which in turn invalidates the inextensibility hypothesis, the inherent non-linear nature of the system may also have a significant contribution to interpret the experimental observations.
In the present work we propose an alternative model for $f$-cones that encompasses the full geometric non-linear contributions, thus capturing any deflections---this model describes the equilibrium shapes as function of the folding (dihedral) angle of the creases. 
Also, the effect of crease mechanics with hingelike behaviors is incorporated into the model. 
Then, we corroborate the predictions of the model with the aid of FEM simulations.
From the numerical model we are able to quantify the stretching on the system in order to test the validity of the inextensibility hypothesis during the entire indentation process.

The manuscript is organized as follows. 
In section II the system under study and its geometry are presented in detail.
Then, in section III, we present our elastic model for $f$-cones and the main results. 
The results presented here complement the predictions of the linear version of the model~\cite{lechenault2015generic}. 
Subsequently, in section IV, a numerical model that simulate an $f$-cone of 4 creases is proposed to study the snapping process in a finite element analysis. 
Then, in section V the results of the numerical study are compared with the theory. 
The details of the analytical calculations can be found in the Appendix.

\section{Kinematical description of nonrigid single vertex origami}

Foldable cones, or $f$-cones, are made from a circular elastic sheet decorated by one or more straight radial creases meeting at a single vertex~\cite{lechenault2015generic}. 
These surfaces resemble those of $d$-cones~\cite{ben1997crumpled}, except that they can fold around the creases. 
When the elastic sheet is inextensible, the only possible equilibrium shapes are developable surfaces and, in this particular geometry, developable cones. 
This implies that the deformed shape can always be isometrically mapped to the initial flat state. 
The equilibrium shape of the cone will be developable anywhere except at the tip of the cone and the creases, where the curvature is not defined. In this section, we first introduce the parametrization of a general conical shape, and then we describe in detail the geometry of an $f$-cone. 

\begin{figure}[tbh]
\centering
\includegraphics[width=0.9\columnwidth]{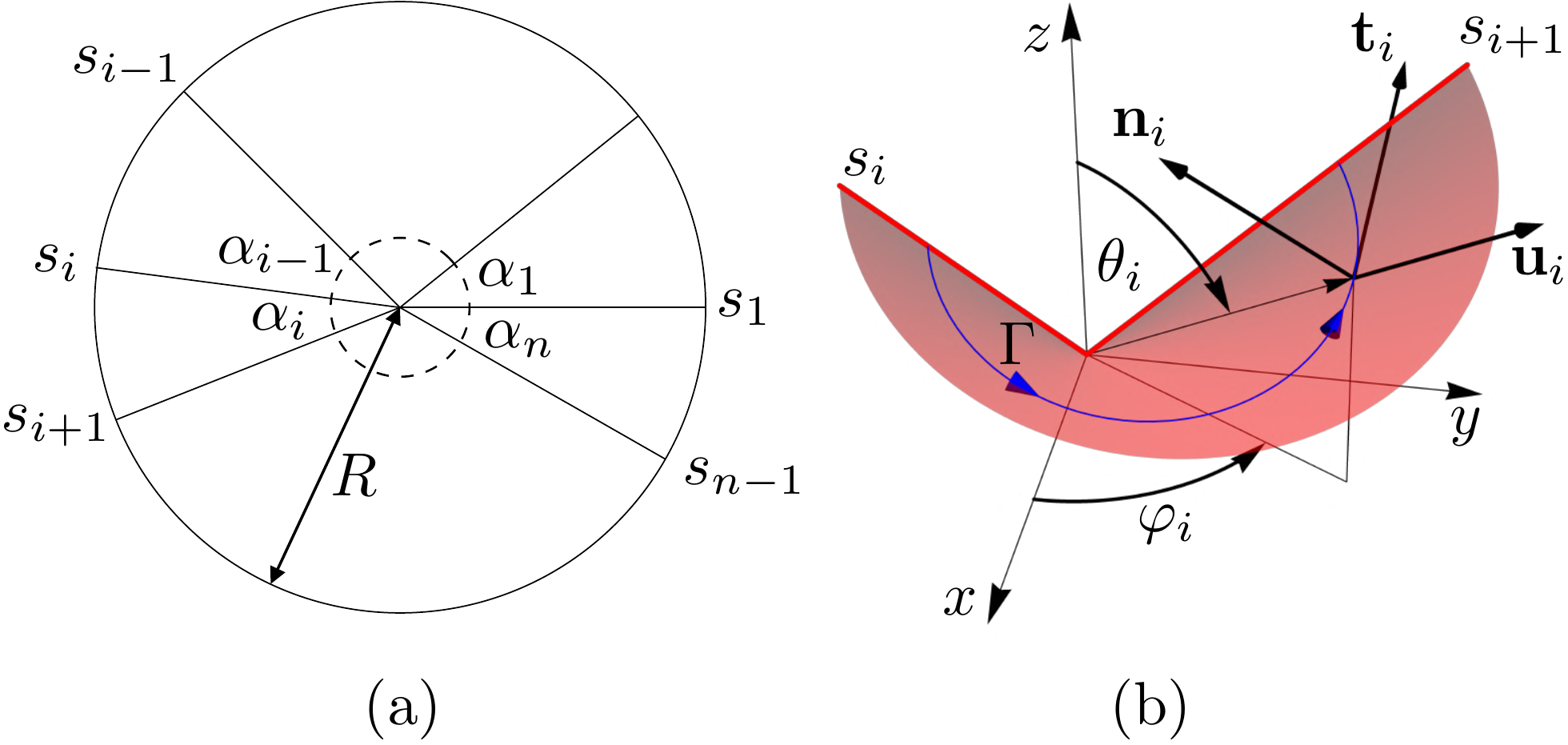}
\caption{(a) Imprinted crease pattern on a flat plate. (b) Deformed state of the $i^{th}$ panel of a $f$-cone. The curve $\Gamma$, the material frame and the Euler-like angles are defined.}
\label{Fig: Panel1}
\end{figure}

\subsection{Geometry of developable cones}

The most general parametrization of a conical shape is given by $\mathbf{r}(r,s) = r \mathbf{u}(s)$,
where $r$ is the distance to the tip, $\mathbf{u} (s)$ is a unit vector and $s \in \left[ 0,2\pi\right]$ is the arc-length of the curve $\Gamma: s \rightarrow ~\mathbf{u}(s)$ on the unit sphere.
The tangent vectors adapted to the surface of the cone are $\mathbf{u}$ and $\mathbf{t}=\mathbf{u}'$, where the prime denotes derivative with respect to $s$. 
As $s$ is the arc-length of the curve, the tangent vector $\mathbf{t}$ is a unit vector. 
Note that $\mathbf{u}\cdot \mathbf{t}=0$ and that the normal of the surface is given by $\mathbf{n}=\mathbf{u}\times \mathbf{t}$. 
Therefore, the triad $\left\lbrace \mathbf{u},\mathbf{t},\mathbf{n}\right\rbrace $ forms a right-handed basis that satisfies the following equations~\cite{cerda2004elements,cerda2005confined}
\begin{subequations}
	\label{eq:Frame}
		\begin{align}
			\mathbf{u}' &= \mathbf{t} 
				\label{Derivada de u}\\*
			\mathbf{t}' &= -\kappa \mathbf{n} - \mathbf{u} 
				\label{Derivada de t}\\
			\mathbf{n}' &= \kappa \mathbf{t}, 
				\label{Derivada de n}
	\end{align}
\end{subequations}	
where $\kappa (s)= \mathbf{t}(s)\cdot \mathbf{n}'(s)$.
The metric tensor of the conical surface is given by $g_{ab} = \partial_{a} \mathbf{r} \cdot \partial_{b} \mathbf{r} $, where the indices $a,b=r,s$. 
Hence, for conical geometries, the metric components are $g_{rr}=1$, $g_{rs}=0$ and $g_{ss}=r^2$. 
The extrinsic curvature tensor is defined as $K_{ab} = \partial_{a} \mathbf{r} \cdot \partial_{b} \mathbf{n}$ and its single non-vanishing component is $K_{ss}=r \kappa$. Therefore, the surface curvature is $K=g^{ab}K_{ab}=\kappa /r$. 
Once $\kappa(s)$ is known, the final shape of the cone can be reconstructed by integrating Eqs.~\eqref{eq:Frame}.

\subsection{Geometry of foldable cones}

We consider a $f$-cone of $n$ creases made from a flat circular sheet of radius $R$ which is parceled out in $n$ circular sectors (panels) delimited by the creases (see Fig.~\ref{Fig: Panel1}(a)). 
A hole of radius $r_0\ll R$ is cut out at the center in order to avoid a divergence in the elastic energy.
Let $\alpha_i$ denote the sector angle of the $i^{th}$ panel in the flat configuration, with $\sum^{n}_{i=1} \alpha_i=~2\pi$. 
The value of the arc-length at the $i^{th}$ crease is denoted by $s_i$, so that, through the inextensibility condition, $\alpha_i=s_{i+1}-s_i$ in the deformed configuration.
Hereinafter, for any scalar or vector field of the form $b_i (s)$, the subscript $i$ specifies that the domain of the function corresponds to the $i$-th sector, where the periodic convention $b_{i\pm n} \equiv b_i$ is assumed.
Moreover, we introduce the following notation: $b^{-}_i\equiv b_{i-1} (s_i)$ and $b^{+}_i\equiv b_i (s_i)$.

\begin{figure}[tbh]
\centering
\includegraphics[width=0.9\columnwidth]{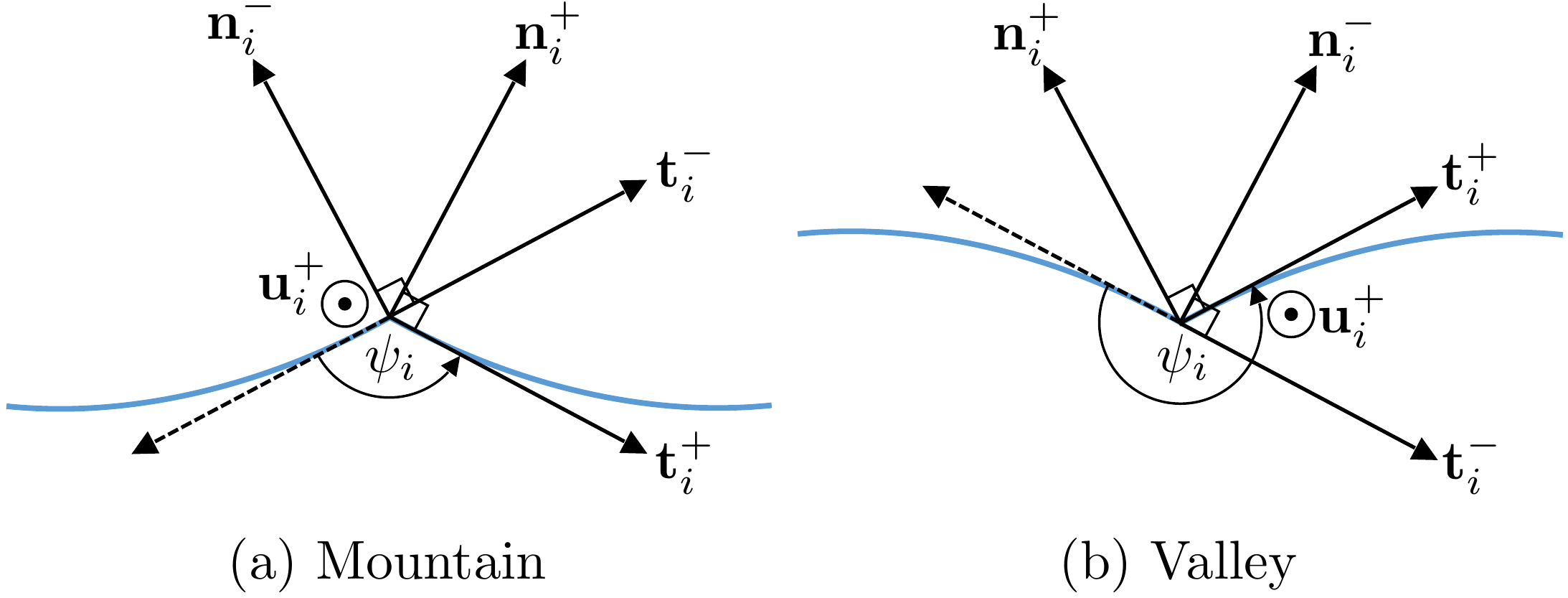}
\caption{Definition of a mountain and a valley creases. The vectors $\mathbf{t}^{\pm}_i$, $\mathbf{n}^{\pm}_i$ and the crease angle $\psi$ are defined.}
\label{Fig: MountainValley}
\end{figure}

In the deformed configuration, each crease has a folding angle (dihedral angle) $\psi_i$, which we call it \textit{mountain} if $\psi_i \in [0,\pi] $ or a \textit{valley} if $\psi_i \in [\pi,2\pi]$. 
For a mountain and a valley creases, Fig.~\ref{Fig: MountainValley} shows how the folding angle is defined by $\mathbf{t}^{+}_i$ and $\mathbf{t}^{-}_i$ in the plane perpendicular to the crease.
In terms of spherical coordinates, the final shape is given by a polar angle $\theta(s)$ and an azimuthal angle $\varphi(s)$ which are functions of the arc-length [Fig.~\ref{Fig: Panel1}(b)].
Each angular sector will span an azimuthal angle $\Delta \varphi_i \equiv \varphi (s_{i+1})-\varphi (s_i)$. The closure condition can be written as
\begin{equation}
\sum^{n}_{i=1} \Delta \varphi_i =\left\{\begin{array}{ll}
\pm 2\pi& \text{if } \Gamma \text{ encloses the }z\text{-axis}\\
0 & \text{if not} 
\end{array}\right.,
\label{eq:closure}
\end{equation}
where $\pm$ indicates that $\varphi_i$ could increase clockwise or counter-clockwise in the $x$-$y$ plane. Notice that in principle $\Delta \varphi_i$ and $\alpha_i$ are not necessarily equal in the deformed configuration. 
However, they coincide in certain symmetrical cases: $f$-cones with an arbitrary number of evenly distributed mountain creases or an even number of evenly distributed alternating mountain-valley creases where all the creases are identical.
In such cases, we shall say in the following sections the $f$-cone is \textit{symmetrical}.

\section{Elastic theory of foldable cones}

Our model is based on a generalization of the functional introduced in Ref.~\cite{guven2008paper}.
The total energy of an $f$-cones with $n$ creases is the sum of the elastic energy over all the panels plus the mechanical energy stored in the creases.
Thus, the principle of virtual work is equivalent to minimizing the following functional
\begin{eqnarray}
F_{n} [\mathbf{u},\mathbf{t}] &= & a\,\sum^{n}_{i=1} \int^{s_{i+1}}_{s_i}  \left[\frac{1}{2} (\mathbf{u}_i\cdot \mathbf{t}_i  \times \mathbf{t}'_i)^2 + \frac{\lambda_i}{2} (\mathbf{u}^2_i -1) \right. \nonumber \\
&& \left. + \frac{\Lambda_i}{2} (\mathbf{t}^2_i -1)+  \mathbf{f}_i \cdot (\mathbf{t}_i -\mathbf{u}'_i) \right]ds \nonumber \\
&& + \sum^{n}_{i=1} g_i \left[\mathbf{t}^{-}_i,\mathbf{t}^{+}_i,\mathbf{u}^{+}_i\right] . 
\label{Functional panels}
\end{eqnarray}
Here,  $a =B \ln \left( R/r_0 \right)$, where $B$ is the flexural stiffness (bending modulus) of the sheet. 
The first term inside the brackets accounts for the bending energy of the facets, where $\mathbf{u}_i(s) \cdot ( \mathbf{t}_i(s) \times \mathbf{t}'_i(s))=\kappa_i(s)$ is the dimensionless curvature of the  $i$-th panel.
The above augmented energy functional contains $3n$ local Lagrange multipliers, namely $\lambda_i(s)$, $\Lambda_i (s)$ and $\mathbf{f}_i(s)$, which correspond to the following kinematical constraints, respectively: $\lambda_i(s)$ enforces $\mathbf{u}_i$ to be a unit vector, thus constraining the final trajectories to the unit sphere; $\Lambda_i (s)$ enforces the parameter $s$ to be the arc-length of the curve $\Gamma$; and finally, $\mathbf{f}(s)$ is a force (normalized by $a$) that anchors the tangent vector to the embedding. 
The functions $g_i$, which depend on the frame vectors at both sides of the crease, account for the elastic energy stored in the $i$-th crease.
For simplicity, we consider point-like creases, although the model can be generalized to extended creases where a crease is a localized regions with a given natural curvature, as shown in reference~\cite{jules2019local}. 
The variation of functional \eqref{Functional panels} yields a set of $n$ ordinary differential equations given by (see Appendix~\ref{Appendix: Functional variation})
\begin{equation}
 \kappa''_i  + \left( 1  + c_i\right) \kappa_i + \frac{\kappa^{3}_i}{2}= 0, \label{Equation for kappa i}
\end{equation}
where $\{c_i\}^n_{i=1}$ is the set of $n$ integration constants.
The above equation describes the equilibrium shapes of the Euler's \emph{Elastica}.
In the present work we assume that all the panels have the same constant, namely, $c_i=c$ for $i=1,\dots,n$.
By comparing with the linear model of $f$-cones, one notices that $-c$ is proportional to the hoop stress $\sigma_{\varphi \varphi}$ in the limit of small deflections, thus, the sign of $c$ dictates whether the structure is in azimuthal compression ($c>0$) or tension ($c<0$)~\cite{lechenault2015generic,cerda2005confined}. 
The hypothesis of equal constants holds provided that there are no external forces acting on the creases introducing additional stresses in different panels, so that $\sigma_{\varphi \varphi}$ is continuous across the panels.
From varying Eq.~\eqref{Functional panels} one also obtains boundary terms that combine with terms coming from the variation of the energy stored in the creases.
These boundary terms give the natural boundary conditions to solve equation \eqref{Equation for kappa i} for each panel.
In absence of external forces, these terms must satisfy
\begin{equation}
a\displaystyle \sum^{n}_{i=1} \left[ -( \mathbf{f}_i \cdot \delta \mathbf{u}_i) + (\kappa_i \mathbf{n}_i \cdot \delta \mathbf{t}_i )\right]\Big|^{s_{i+1}}_{s_i}+ \displaystyle \sum^{n}_{i=1} \delta g_i  =0,
\label{Eq: Boundary terms + Crease energy}
\end{equation}
where $\mathbf{f}_i = \kappa'_i \mathbf{n}_i - \left(\kappa^{2}_i/2 + c_i \right)\mathbf{t}_i$, which can be interpreted as a normalized force per unit-length along a ray of fixed $r$~\cite{guven2008paper}. 

At this point, it is convenient to introduce the vector $\mathbf{J} \equiv  -\mathbf{u}\times \mathbf{f} +  \kappa \mathbf{u} $ which is a conserved quantity associated with the rotational invariance of the system and can be interpreted as a torque~\cite{guven2008paper}.
It can be shown that the quantity $J^2-c^2$ corresponds to the first integral of equation \eqref{Equation for kappa i}.
One can use the vector $\mathbf{J}$ to obtain the equilibrium shape of the $f$-cone by first setting it parallel to the $z$-axis and projecting it onto the frame $(\mathbf{u},\mathbf{n})$, obtaining $\mathbf{J}\cdot \mathbf{u} = J \cos \theta =  \kappa$ and $\mathbf{J}\cdot \mathbf{n} = J \varphi' \sin^2 \theta =  \kappa^2/2 + c$.

\subsection{Infinitely stiff creases}

In this section, we solve the case of infinitely stiff creases, so that $\delta g_i=0$. 
This means that the set of folding angles $\{\psi_i\}^n_{i=1}$ is an input of the problem and that the final solutions are parameterized by these angles.
In Appendix~\ref{Appendix: Boundary Conditions}, we show that the boundary terms \eqref{Eq: Boundary terms + Crease energy}, together with the condition $\delta \psi_i=0$, imply the following:
\begin{equation}
\kappa^{+}_i=\kappa^{-}_i, \label{Eq: Curvature continuity}
\end{equation}
which means that the curvature is continuous through the crease.
Also, the transversal force $\mathbf{f}$ is continuous, which can be written as
\begin{equation}
\mathbf{f}^{+}_i =\mathbf{f}^{-}_i. \label{Eq: Continuity f}
\end{equation}
The continuity conditions~(\ref{Eq: Curvature continuity},\ref{Eq: Continuity f}) imply that $\mathbf{J}^{+}_i =~ \mathbf{J}^{-}_i$, thus, the entire structure is characterized by a single vector $\mathbf{J}$.

Solving the system \eqref{Equation for kappa i} with the assumption of equal constants, $c_i=c$, requires $2n+1$ boundary conditions.
Combining  equations \eqref{Eq: Curvature continuity} and \eqref{Eq: Continuity f}, one can show that 
\begin{eqnarray}
 \kappa^{\prime+}_i &=& -\kappa^{\prime-}_i, \label{Eq: Condition prime}
\\
  \kappa^{\prime +}_i  &=& - \cot \left(\frac{\psi_i}{2}\right) \left(\frac{{\kappa^{+}_i}^{2}}{2} + c \right), \label{Eq: Condition psi}
\end{eqnarray}
thus, yielding $2n$ boundary conditions.
Adding the closure condition~(\ref{eq:closure}) makes the problem well-posed.

Integration of Eq.~\eqref{Equation for kappa i} for each panel gives two possible solutions:
\begin{equation}
\kappa_i(s) = \left \{
  \begin{aligned}
    &\kappa_{0i} \, \text{cn} \left( \frac{\kappa_{0i}}{2 \sqrt{m^{\mbox{\tiny R}}_i}} s - S_{0i}  \Bigg|  m^{\mbox{\tiny R}}_i \right), && \text{if}\ J^2>c^2, \\
    &\kappa_{0i} \, \text{dn} \left( \frac{ \kappa_{0i}}{2} s - S_{0i} \Bigg|  m^{\mbox{\tiny S}}_i \right), && \text{if}\ J^2<c^2,
  \end{aligned} \right.
  \label{eq:abram}
\end{equation} 
where $ \text{cn} (\cdot | \cdot)$  and $\text{dn} (\cdot | \cdot)$ are the cosine and delta amplitude Jacobian elliptic functions~\cite{abramowitz1965handbook}, with parameters $ m^{\mbox{\tiny R}}_i$ and $m^{\mbox{\tiny S}}_i$ given by
\begin{equation}
 m^{\mbox{\tiny R}}_i= \frac{1}{m^{\mbox{\tiny S}}_i}= \frac{\kappa^2_{0i}}{2 \kappa^{2}_{0i}  + 4 (1 + c)}.
 \label{eq:parameters}
\end{equation}
Here, $\kappa_{0i}$, $S_{0i}$ and $c$ are $2n+1$ unknown parameters that must be fixed such that the boundary conditions and the closure conditions are satisfied.
Without loss of generality, one can define the cosine and delta amplitude functions such that $0<m^{\mbox{\tiny S}}_i<1$ and $0<m^{\mbox{\tiny R}}_i<1$~\cite{abramowitz1965handbook}. Therefore, Eq.~(\ref{eq:parameters}) shows that there exists two families of solutions. 
We identify as the rest configuration (thus the superscript $\mbox{R}$) the solutions for which $J^2>c^2$ and the snapped configuration (superscript $\mbox{S}$) with $J^2<c^2$.
This choice is in agreement with the fact that we find a posteriori that the bending energy of the rest configuration is the lower one.
As the quantity $J^2-c^2$ is defined for the entire structure, all the panels will be in either a rest state, or a snapped state, therefore, there is no mixture of states (provided that there is a single constant $c$).
Notice that the present nonlinear approach allows us to justify the existence of two families of solutions for a given set of folding angles, regardless of whether the cone is symmetrical or not, a property that is difficult to prove in the linear model.

From this point forward, we consider a simplified situation of symmetrical $f$-cones made of all-mountain creases with same folding angle $\psi$ (we shall also omit the subscripts $i$ labelling the panels). 
Thus, it is sufficient to solve Eq.~\eqref{Equation for kappa i} in a single panel where $-\alpha/2 \leq s \leq \alpha/2$ and $\alpha=2 \pi/n$.
For both rest and snapped states, we proceed to find numerically the parameters $\kappa_0$, $S_0$ and $c$ that satisfy the closure condition~(\ref{eq:closure}) and the boundary conditions (\ref{Eq: Condition prime},\ref{Eq: Condition psi}).
The closure condition~(\ref{eq:closure}) reads now $\Delta \varphi= \pm \alpha$ or $0$.
For each state, the quantity $\Delta \varphi$ is an integral that can be computed analytically (see Appendix~\ref{Appendix: Closure condition}).
Then, for a given $S_0$, the closure condition defines a curve in the parameter space $\lbrace \kappa_{0}, c \rbrace$ for each state.
As the $f$-cone is symmetrical, the solutions given by Eq.~(\ref{eq:abram}) should be even functions with respect to $s=0$.
Therefore, one has $S_0=0$ for the rest state and, for the snapped state there are two possibilies, either $S_0=0$ or $S_0=K(m_s)$ ($K(\cdot)$ is the complete elliptic integral of the first kind and corresponds to the half period of the function $\text{dn} (\cdot | m)$)~\cite{abramowitz1965handbook}. 
We only found solutions for $S_{0} = K(m_{s})$ that satisfy the closure condition.
Eq.~(\ref{Eq: Condition prime}) is automatically satisfied by the symmetry of the solutions.
Then, Eq.~\eqref{Eq: Condition psi} defines a second curve in the parameter space $\lbrace \kappa_{0}, c \rbrace$ (one for each state).
Hence, the solution for a given folding angle $\psi$ corresponds to the intersection between these two curves, so that the values of $\kappa_{0} (\psi)$ and $c (\psi)$ are obtained. 
This procedure is done for all $\psi \in [0,\pi]$ ($\psi>\pi$ would correspond to equivalent states but vertically flipped). 

\begin{figure}[tbh]
\centering
\includegraphics[width=\columnwidth]{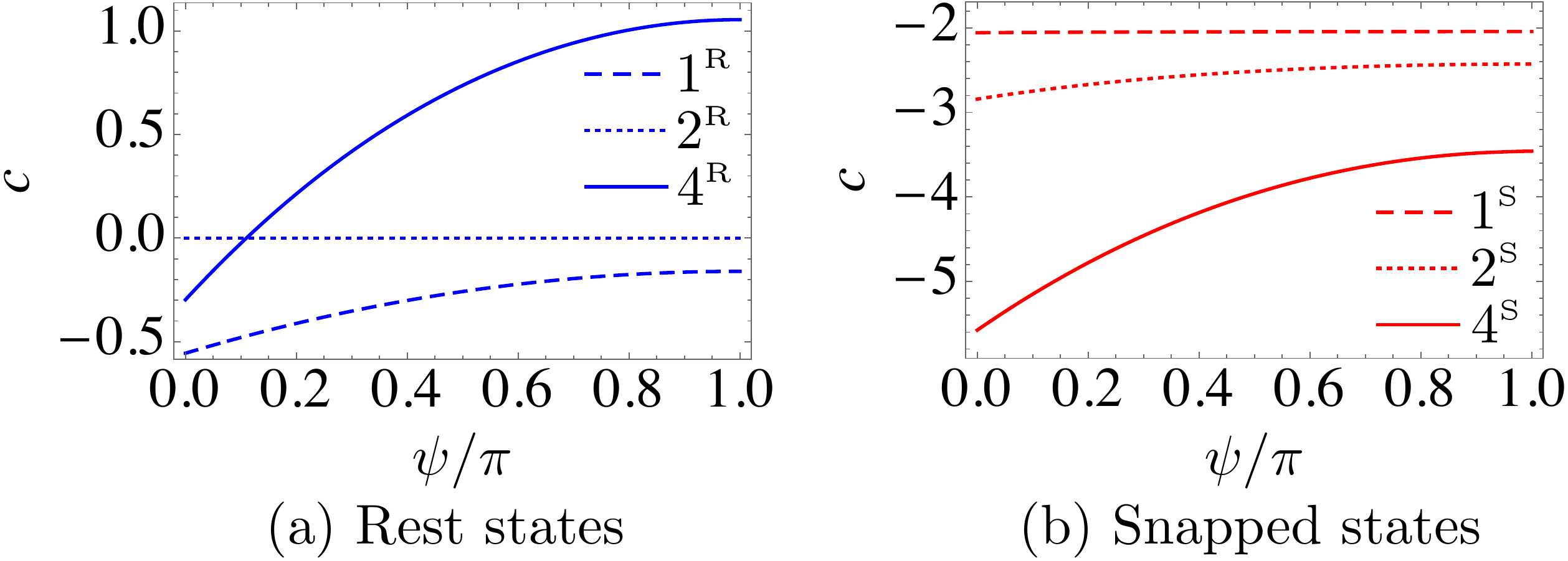}
\caption{The Lagrange multiplier $c$ as function of the folding angle $\psi$ for (a) rest states and (b) snapped states for all-mountain $f$-cones with 1, 2 and 4 creases.}
\label{Fig: Constants}
\end{figure}

Hereinafter, we denote as $n^{\mbox{\tiny R}}$ (resp., $n^{\mbox{\tiny S}}$) the all-mountain $f$-cones with $n$-creases in a rest (resp., snapped) state.
We show here the solutions for the most relevant cases: a semi-infinite crease ($1^{\mbox{\tiny R},\mbox{\tiny S}}$), a single infinite crease ($2^{\mbox{\tiny R},\mbox{\tiny S}}$) and two perpendicular mountain creases ($4^{\mbox{\tiny R},\mbox{\tiny S}}$).
Fig.~\ref{Fig: Constants} shows the resulting $c(\psi)$ for these different cases. 
We noticed that except for the trivial case $2^{\mbox{\tiny R}}$, one has $c\nrightarrow0$ as $\psi \rightarrow \pi$, suggesting the existence of a residual hoop stress as one approaches the flat state.
Indeed, the residual hoop stress at $\psi = \pi$ coincides with the values of the hoop stress in the linear model~\cite{lechenault2015generic}.
We attribute this residual stress to a critical load needed to observe buckling of the facets, as it happens in Euler-Bernoulli beam buckling.
It has been found that the Lagrange multiplier associated with the hoop stress has a fixed value for each state. The Lagrange multiplier $c$ depends on the folding angle, except for $1^{\mbox{\tiny S}}$ where $c(\psi)$ is nearly constant. 
In the configuration $4^{\mbox{\tiny R}}$, $c$ changes sign as $\psi \rightarrow 0$, suggesting at first sight a modification of the hoop stress from compressive to tensile as the folding angle gets sharper. 
However, this could be misleading because the interpretation of $c$ as a hoop stress is valid only for small deformations.

\begin{figure}[tbh]
\centering
\includegraphics[width=\columnwidth]{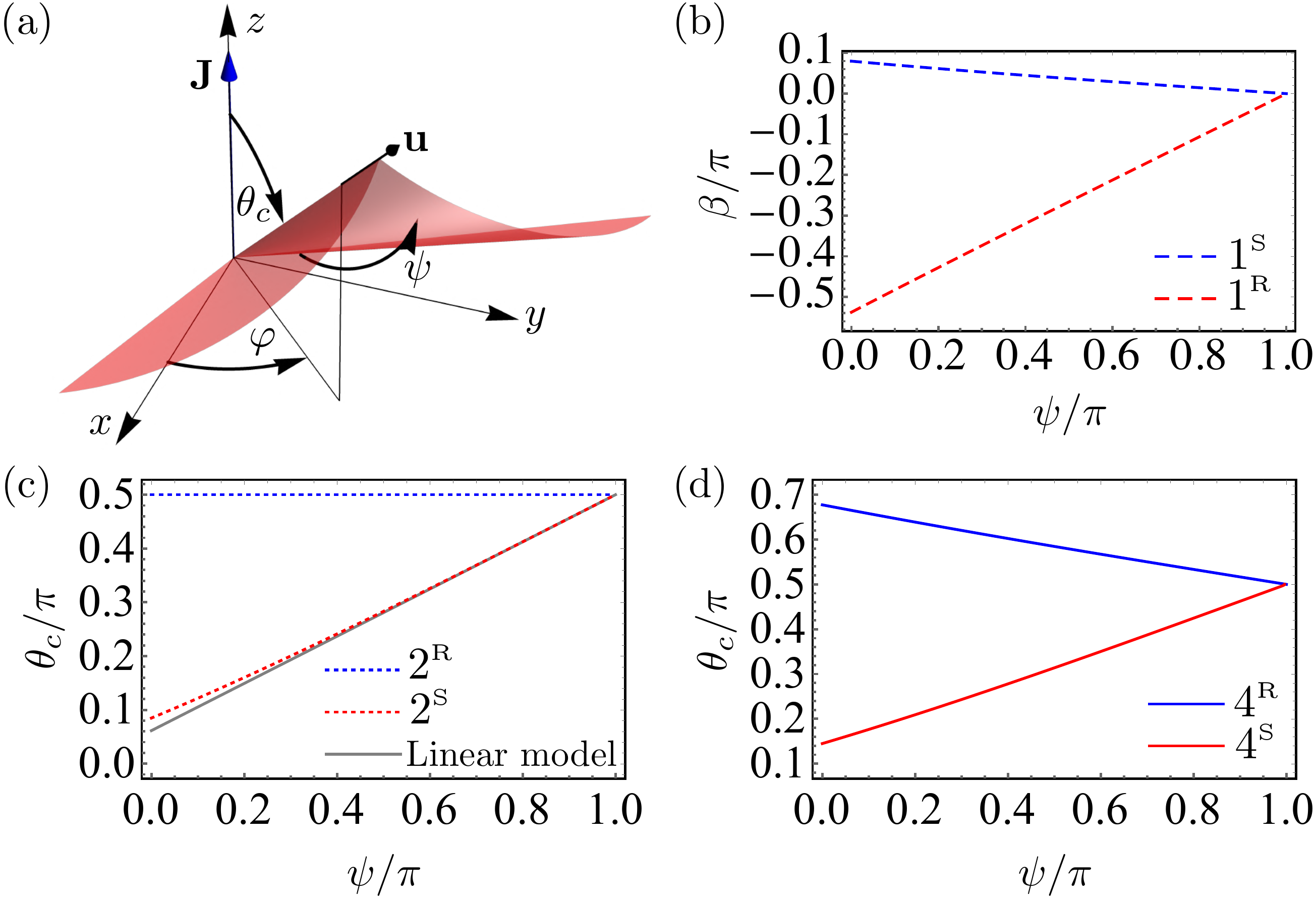}
\caption{(a) Schematic definition of the polar angle $\theta_c$ for $n^{\mbox{\tiny R},\mbox{\tiny S}}$ $f$-cones ($n>1$).
(b) Case $1^{\mbox{\tiny R},\mbox{\tiny S}}$: the angle $\beta(\psi)=\pi-\cos^{-1} (\mathbf{u}(0)\cdot\mathbf{u}(\pi))$ that describes the deviation from the flat configuration.
(c) Case $2^{\mbox{\tiny R},\mbox{\tiny S}}$: the polar angle $\theta_{c}(\psi)$ at the creases compared with the prediction of the linear model for the snapped state.
(d) Case $4^{\mbox{\tiny R},\mbox{\tiny S}}$: the polar angle $\theta_{c}(\psi)$ at the creases.
}
\label{Fig: Polar angles}
\end{figure}

The equilibrium shapes of the $f$-cones are plotted using the coordinates $\theta(s)$, $\varphi(s)$.
Thus, they are universal in the sense that they depend only on the folding angle $\psi$ and not the material properties.
Nevertheless, changing the bending modulus or the dimensions of the cone will modify the stresses and torques supported by the system as well as its elastic response.
Fig.~\ref{Fig: Polar angles} shows the deviations of the structure from the flat state. 
To quantify such deviations, we use the angle $\beta=\pi-\cos^{-1} (\mathbf{u}(0)\cdot\mathbf{u}(\pi))$ for a $1^{\mbox{\tiny R},\mbox{\tiny S}}$ $f$-cone and the polar angle $\theta_c= \cos^{-1}(\mathbf{z}\cdot\mathbf{u}(\alpha/2))$ for $n^{\mbox{\tiny R},\mbox{\tiny S}}$ $f$-cones ($n>1$). 
The choice of $\psi$, instead of $\theta_{c}$ (or $\beta$ for a $1^{\mbox{\tiny R},\mbox{\tiny S}}$ $f$-cone) as a control parameter prevents finding unphysical self-intersecting solutions.
In Fig.~\ref{Fig: Polar angles}(c), we compare $\theta_{c}(\psi)$ for $2^{\mbox{\tiny S}}$ configuration with the prediction of the linear model $\theta_{c}=\pi/2+0.4386(\psi-\pi)$~\cite{lechenault2015generic,walker2018shape}.
The polar angle of the crease deviates slightly from the linear prediction for sharper folding angles.
This deviation fits better the experimental results shown in reference~\cite{walker2018shape}, although there is still an important discrepancy from the theory because of the limits of the inextensibility hypothesis, which we explore in more detail in section~IV.

\subsection{Crease mechanics}

The mechanical energy of a single point-like crease can be written as a function of invariants built from the three unit vectors that define the crease geometry: the crease vector and two vectors tangent to each facet~\cite{brunck2016elastic}.
Such invariants are $ \mathbf{t}^{-}_i \cdot \mathbf{t}^{+}_i $ and $\left(\mathbf{t}^{-}_i \times \mathbf{t}^{+}_i  \right) \cdot \mathbf{u}^{+}_i$.
At leading order, the elastic energy of the $i$-th crease takes the form~\cite{brunck2016elastic}
\begin{equation}
g_i = L \, \left(\sigma_i \, \mathbf{t}^{-}_i \cdot \mathbf{t}^{+}_i   + \tau_i \left(\mathbf{t}^{-}_i \times \mathbf{t}^{+}_i  \right) \cdot \mathbf{u}^{+}_i\right), \label{Crease energy}
\end{equation}
where $L=R-r_0$, $\sigma_i$ and $\tau_i$ are material constants associated to the crease. 
The crease energy can be rewritten in terms of the folding angle $\psi_i$, defined as the oriented angle $(\widehat{-\mathbf{t}^{-}_i,\mathbf{t}^{+}_i} )$ (see Fig.~\ref{Fig: MountainValley}). 
Introducing the constants $k_i$ and $\psi^{0}_i$, such that
\begin{equation}
\sigma_i = k_i \cos \, \psi^{0}_i, \hspace{5mm}  \tau_i = k_i \sin \, \psi^{0}_i,  \label{Crease constants}
\end{equation}
allows us to rewrite \eqref{Crease energy} as $g_i (\psi_i) = - L \, k_i \, \cos \left( \psi_i - \psi^{0}_i \right)$. Thus, the crease energy $g_i = g_i (\psi_i)$ is a function of the folding angle which is now an unknown variable. 
If $\psi_i \approx \psi^{0}_i $, the crease energy approximates to the energy of an elastic hinge $g_i \approx~L \, k_i \, (\psi_i - \psi^{0}_i)^2 /2 + E_0$,
where $E_{0}$ is a constant, $k_i$ is the crease stiffness, and $\psi^{0}_i$ is the rest angle of the crease. 
The conical geometry implies that the folding angle is constant along the crease.
One can show that taking into account the terms coming from the variation of the crease energy, we obtain an additional boundary condition (see Appendix~\ref{Appendix: Boundary Conditions})
\begin{equation}
\kappa^{+}_i= \frac{1}{a} \frac{d g_i}{d \psi_i}. \label{Crease energy boundary condition}
 \end{equation}
Eq.~(\ref{Crease energy boundary condition}) states that the value of the curvature at the crease is given by the moment imposed by its mechanical response.
For a crease energy given by Eq.~(\ref{Crease energy}), one has $\kappa^{+}_i= \bar{k}_i \, \sin \left( \psi_i - \psi^{0}_i \right)$, where $\bar{k}_i =L k_i /a$ is  the normalized crease stiffness.

We study again configurations with equally spaced mountain creases that have the same mechanical properties, such that, $\bar{k}_i=\bar{k}$ and $\psi_i^{0}=\psi_{0}$ .
The control parameters are then the crease stiffness $\bar{k}$ and the rest angle $\psi_{0}$. 
The boundary conditions are those of the infinitely stiff creases case supplemented by Eq.~(\ref{Crease energy boundary condition}). Therefore, one can use the solutions found for the infinitely stiff creases and search the value of $\psi$ such that Eq.~\eqref{Crease energy boundary condition} is satisfied.

\begin{figure}[tbh]
\centering
\includegraphics[width=\columnwidth]{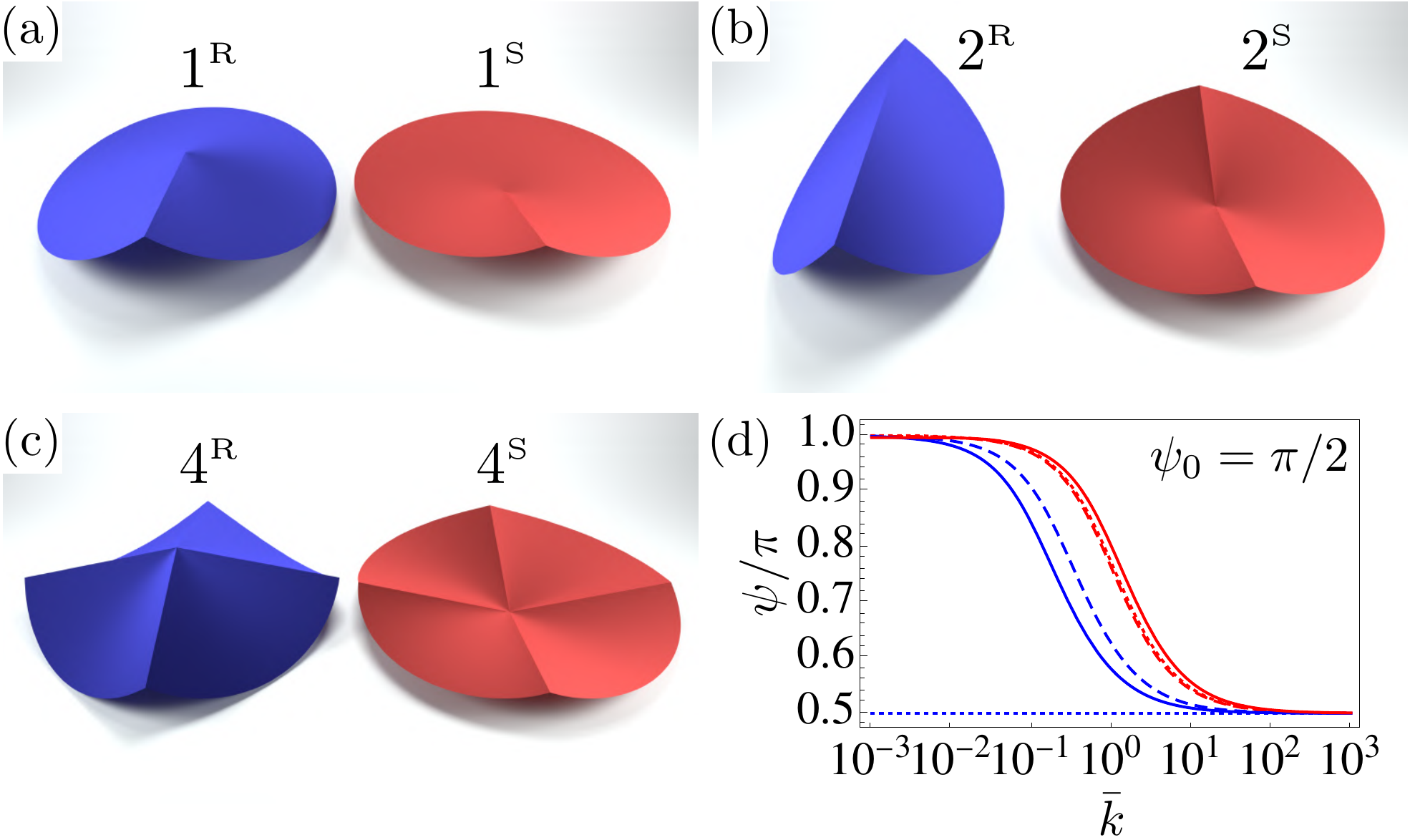}
\caption{(a-c) Equilibrium shapes for $\bar{k}=1$ and $\psi_{0}=\pi/2$. (d) Final crease angles $\psi$ as function of $\bar{k}$ for $\psi_{0}=\pi/2$. Rest (blue lines) and snapped (red lines) states for all-mountain $f$-cones with 1, 2 and 4 creases (respectively dashed, dotted and plain lines).}
\label{Fig: Equilibrium shapes}
\end{figure}

Figures~\ref{Fig: Equilibrium shapes}(a-c) show typical shapes for the cases $1^{\mbox{\tiny R},\mbox{\tiny S}},2^{\mbox{\tiny R},\mbox{\tiny S}},4^{\mbox{\tiny R},\mbox{\tiny S}}$.
Fig.~\ref{Fig: Equilibrium shapes}(d) shows the final folding angle $\psi$ as function of $\bar{k}$ for a fixed rest angle $\psi_{0}$ of the crease.
Notice that $\psi \rightarrow \pi$ as $\bar{k} \rightarrow 0$ and that $\psi \rightarrow \psi_{0}$ as $\bar{k} \rightarrow \infty$.
For a given $\bar{k}$, the snapped state always displays a larger $\psi$ than the rest state.
This observation is explained by the fact that the hoop stress of the snapped state is always larger than its respective rest state.

\begin{figure}[tbh]
\centering
\includegraphics[width=\columnwidth]{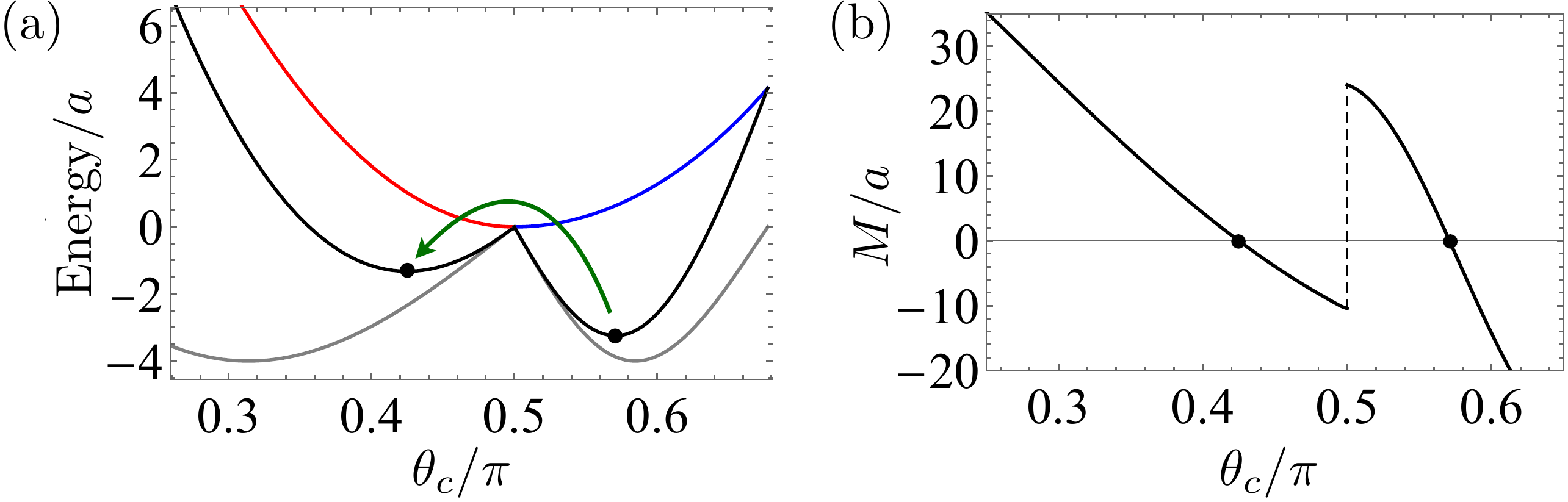}
\caption{
(a) Energy landscape of an all-mountain $f$-cone with 4 creases as function of $\theta_c$ for $\bar{\kappa}=1$ and $\psi_0=\pi/2$. 
The blue and red curves correspond to the bending energy of the facets for the rest and snapped states, respectively. 
The plain gray curve is the crease energy given by Eq.~\eqref{Crease energy}. 
The black curve is the total elastic energy.
(b) Normalized moment $M(\theta_c)$ applied on the creases. The black dots correspond to the rest and snapped states of the $f$-cone.}
\label{Fig: Energy Landscape}
\end{figure}

The total energy of the structure can be computed by summing the bending energy of the facets and the energy of the creases.
Fig.~\ref{Fig: Energy Landscape}(a) shows an example of the energy landscape as function of the polar angle $\theta_c$ of the crease for an all-mountain $f$-cone with 4 creases.
While the bending energy of the facets has an asymmetric parabolic shape, the energy of the creases has a double-well potential shape whose minima are at the same energy level. 
The resulting shape of the total energy is an asymmetric double-well potential where the two minima corresponds to the rest and the snap states. 
Figure~\ref{Fig: Energy Landscape}~(b) shows the derivative of the energy with respect to $\theta_c$ which corresponds to the moment $M(\theta_c)$ applied on the creases.
Notice that the energy has a cusp at $\theta_c=\pi/2$ which leads to a discontinuity in the mechanical response of the structure.
This type of snap-through transition has been found in similar systems such as the waterbomb origami with rigid facets~\cite{hanna2014waterbomb}.
However, the snap-through mechanisms of the $f$-cone and the waterbomb are different. The former is mediated by the asymmetry in the bending energy between the rest and snapped states and the latter by the asymmetrical kinematical conditions imposed by the rigid facets.
For the waterbomb, the only relevant energy is the mechanical energy stored in the creases which, in addition to the kinematical constraints, accounts for an asymmetric double-well like potential.

\section{Continuous elastic model of origami structures}

Commonly, the mechanical response of origami-based metamaterials can not be reduced to elastic hinges connected to rigid or isometric panels~\cite{silverberg2015origami}. The energy landscape of deformations depends generally on both bending and stretching energies of the panels as well as on the inherent spatially extended nature of the creases~\cite{jules2019local}. Therefore, one needs to supplement the present analysis with a more accurate description that takes into account these different contributions. When applied to the $f$-cone, such a description should be validated by the bounds given by the analytical model.

In the following, we propose a mechanical model based on a continuous description of origami structures that is suitable for numerical implementation
and test it on the $f$-cone by performing  Finite Element analysis (FEA).
The simulations were designed in the commercial FEA software COMSOL Multiphysics 5.4. 
Within this software, the Structural Mechanics Module is equipped with quadratic shell elements, which have been used to our purpose. 
All the simulations were carried out with a linear elastic Hookean material model and geometric non-linear kinematic relations have been included. 
The plate Young's modulus is $E=3.5$ GPa, the Poisson's ratio is $\nu=0.39$, and the plate thickness is $h=300 \, \mu$m.
We searched for solutions with the default stationary solver, where the non-linear Newton method has been implemented. Mesh refinement studies were undertaken to ensure convergence of the results.

\subsection{Temperature-induced hingelike creases}

\begin{figure}[tbh]
\centering
\includegraphics[width=\columnwidth]{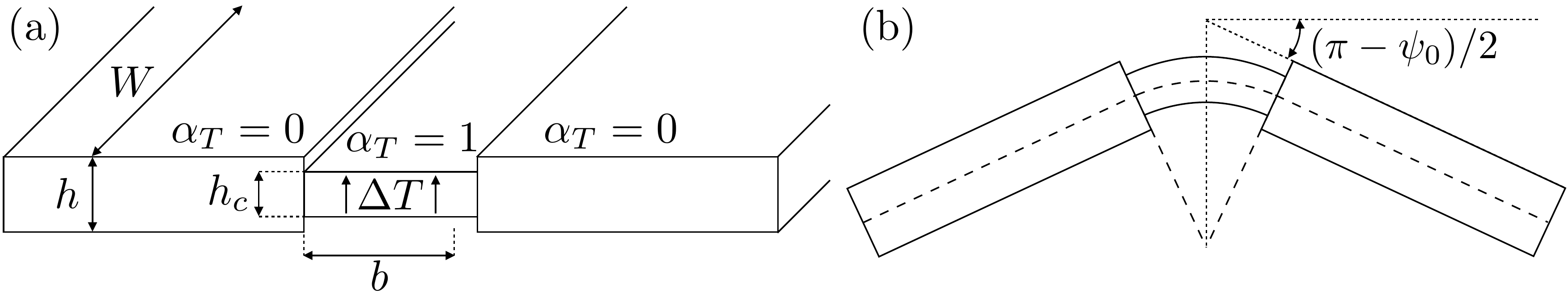}
\caption{Transverse view of a plate with a narrow slice whose thermal and mechanical properties differ from those of the rest of the plate. (a) Reference configuration. (b) Curvature induced equilibrium configuration due to a linear thermal gradient across the thickness.}
\label{Fig: Crease Diagram}
\end{figure}

In our numerical model, we develop a method to create creases that are able to reproduce the hingelike mechanical response of commonly folded thin sheets. 
Inspired by the experimental results of Ref.~\cite{jules2019local}, we model creases as narrow slices of the plate undergoing thermal expansion due to a temperature gradient through the thickness of the plate, as it is schematically shown in Fig.~\ref{Fig: Crease Diagram}.  

In order to test the mechanical response of these temperature driven creases, we first perform a numerical test of a single fold in a hingelike geometry consisting of two facets and the crease in the middle. 
We consider a rectangular plate of length $L$, width $W$ and thickness $h$.
In the middle of the plate, we take a transversal narrow slice of width $b \ll W$, across the width of the plate, dividing the plate into three parts. 
The narrow slice corresponds to the crease while the other two parts correspond to the facets. 
A linear temperature gradient $\Delta T$ is applied through the thickness of the plate. 
We let the creased region of the plate to undergo thermal expansion by defining an inhomogeneous coefficient of thermal expansion $\alpha_{T} (s)$ which is constant for $|s|<b$ and zero elsewhere, where $s$ is the arc-length perpendicular to the creases.
To simulate a more realistic crease, we add a rigid connector made of an infinitely stiff material of length $W$, thus preventing bending deformation in the direction of the crease line. 
Moreover, tuning the mechanical response of the crease is done by varying the thickness $h_c$ and the Young's modulus $E_c$ of the crease slice.
The parameters $h_c$, $E_c$, $\Delta T$ and $b$ will define the crease mechanical response.

Taking advantage of the two-plane symmetry of the single fold geometry, we solve only for one quarter of the plate and then obtain the entire equilibrium shape by reflections.
We perform two different studies: a heating up test and a mechanical response test.
The first study consists in heating the plate from the bottom while the ends of the two plates parallel to the crease are constrained to move in the $xy$-plane.
When heating, the crease bends towards the sense of lower temperature (bottom), while the facets remain practically flat. 
The resulting angle between the two facets corresponds to the rest angle $\psi_0$ of the crease.
The second study consists in a test of the mechanical response of the crease. 
We add four additional rigid connectors to the sides of the facets that are perpendicular to the fold line.
Two moments of opposite signs are applied respectively to each pair of rigid connectors attached to the facets so that the hingelike system can open or close.
Through this mechanical test, we were able to verify the hingelike behavior of the crease. 

In the following, we employ the temperature-induced creases in our numerical model of $f$-cones.

\section{Numerical analysis of foldable cones}

We begin with a circular planar disc of external radius $R=100$ mm and a central hole of radius $r_0=1$ mm.
Then, $n$ radial narrow slices of constant width $b$, that correspond to the creases of the $f$-cone, are created.
Rigid connectors along the creases are added so as to prevent bending along the longitudinal direction of the creases.
In order to take advantage of the symmetries of the system, depending on the number of creases, only the fundamental unit cells are numerically solved and then the complete structure is reconstructed using reflections through the symmetry planes.
Because each plane of symmetry cannot coincide with a rigid connector, the $1^{\mbox{\tiny R},\mbox{\tiny S}}$ case must be solved entirely.
For $2^{\mbox{\tiny R},\mbox{\tiny S}}$, only a half of the disk is numerically solved so that the axis of symmetry is perpendicular to both creases.
For $4^{\mbox{\tiny R},\mbox{\tiny S}}$, a quarter of the disk is solved so that a single crease is at $45º$ from one plane of symmetry.

\subsection{Indentation tests} 

Hereinafter, we focus only in the all-mountain $f$-cone with 4 creases, anticipating that our general conclusions also apply to more complex configurations.
In order to test our analytical predictions, we study the snapping of the system through an indentation process from the rest state to the snapped state, which is carried out in two steps.
Initially, we turn on the temperature to take the $f$-cone to its rest state with a given folding angle $\psi$.
The ends of the creases are constrained to move in the $xy$-plane, so that the points at the central rim rise up when the temperature is activated.
For all our simulations, we fix the temperature gradient in the crease such that $\alpha_T \Delta T = 0.2$.
We also choose $b=1$~mm, which is within the expected order of magnitude with respect to $h$ according to crease formation measurements in thin sheets~\cite{benusiglio2012anatomy}.

In a second step, the central rim is vertically lowered quasistatically while constraining the ends of the creases to move in the $xy$-plane.
Each crease is constrained to rotate only in the plane defined by its initial direction and the $z$-axis (as shown in Fig.~\ref{Fig: Polar angles}(a)).
Throughout indentation, the vertical displacement of the central rim is specified and a reaction moment $M(\theta_c)$ at the creases is computed.
Therefore, the mechanical response of the $f$-cone to the indentation process consists in the determination of the curve $M(\theta_c)$ (see movies available as supplemental material of such indentation tests in~\cite{Supp}).

The folding angle $\psi$ of the creases is also tracked during indentation. To measure $\psi$ from the numerical results, we extract the tangent vector field of concentric curves initially defined in the flat configuration.
We evaluate this vector field at each side of the crease and measure the resulting angle between them. 
The local folding angle is found to be not exactly constant along the crease but is a function of the radial coordinate. 
For this reason, a representative measurement of $\psi$ is chosen to be the average between two local folding angles measured at radial distances $r_0+(R-r_0)/3$ and $r_0+2(R-r_0)/3)$. 
In each indentation test, we extract a curve $\theta_c$ as function of $\psi$ which we call the indentation path.

The resulting curves $M(\theta_c)$ and $\theta_c(\psi)$ will be discussed in the following.

\subsection{Numerical results}

\newcommand{\colorfourline}{\raisebox{2pt}{\tikz{\draw[-,color4,solid,line width = 0.9pt](0,0) -- (5mm,0);}}}
\newcommand{\colortenline}{\raisebox{2pt}{\tikz{\draw[-,color10,solid,line width = 0.9pt](0,0) -- (5mm,0);}}}
\begin{figure}[tbh]
\centering
\includegraphics[width=\columnwidth]{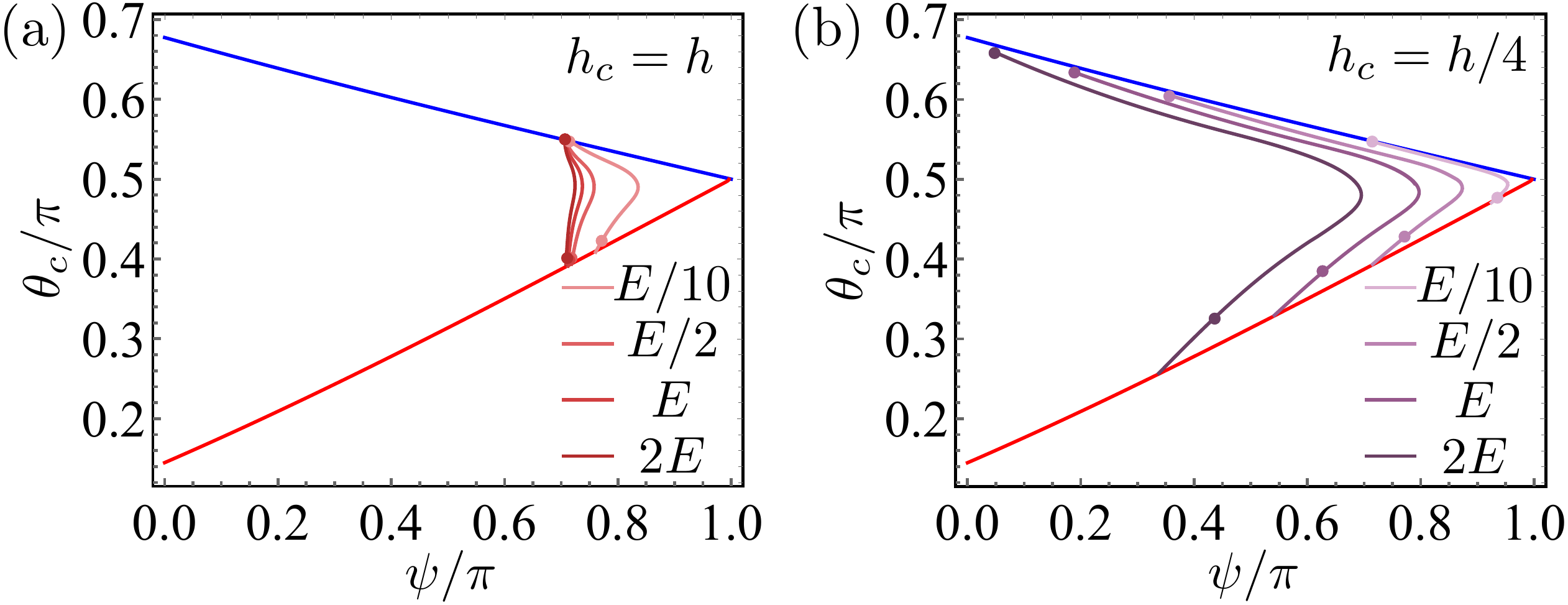}
\caption{(a-b) Indentation paths $\theta_c(\psi)$ as computed by the numerical model for $h_c/h=1,1/4$ and  $E_c/E=2,1,1/2,1/10$. The theoretical curves $\theta_c(\psi)$ for a $4^{\mbox{\tiny R},\mbox{\tiny S}}$ $f$-cone are reproduced from Fig.~\ref{Fig: Polar angles}. The dots correspond to the rest and snapped states of each indentation path.}
\label{Fig: Indentation Paths}
\end{figure}

Fig.~\ref{Fig: Indentation Paths} shows the variation of $\theta_c(\psi)$ obtained numerically during the indentation test for different crease thicknesses and Young moduli. These results are compared with the polar angle $\theta_{c}(\psi)$ given by analytical $f$-cone calculations shown in Fig.~\ref{Fig: Polar angles}(d). 
This parametric study allows us to highlight to what extent the softness of the crease affects the indentation path. Our results show that our continuous model of the crease is more sensitive to variations of $h_c$ than those of $E_c$.  We notice that the indentation paths do not generally follow the analytical solutions given by the isometric constraint, meaning that the intermediate shapes throughout indentation are not perfect developable cones. 
If the crease is too stiff, as in the case of Fig.~\ref{Fig: Indentation Paths}(a), the indentation path follows a nearly vertical line (\emph{i.e.} approximately constant folding angle path) connecting the two stable points. However, when the crease is made softer (Fig.~\ref{Fig: Indentation Paths}(b)), either by reducing its thickness or its Young's modulus, the indentation paths approach the one predicted by the isometric constraint.

\begin{figure}[tbh]
\centering
\includegraphics[width=.9\columnwidth]{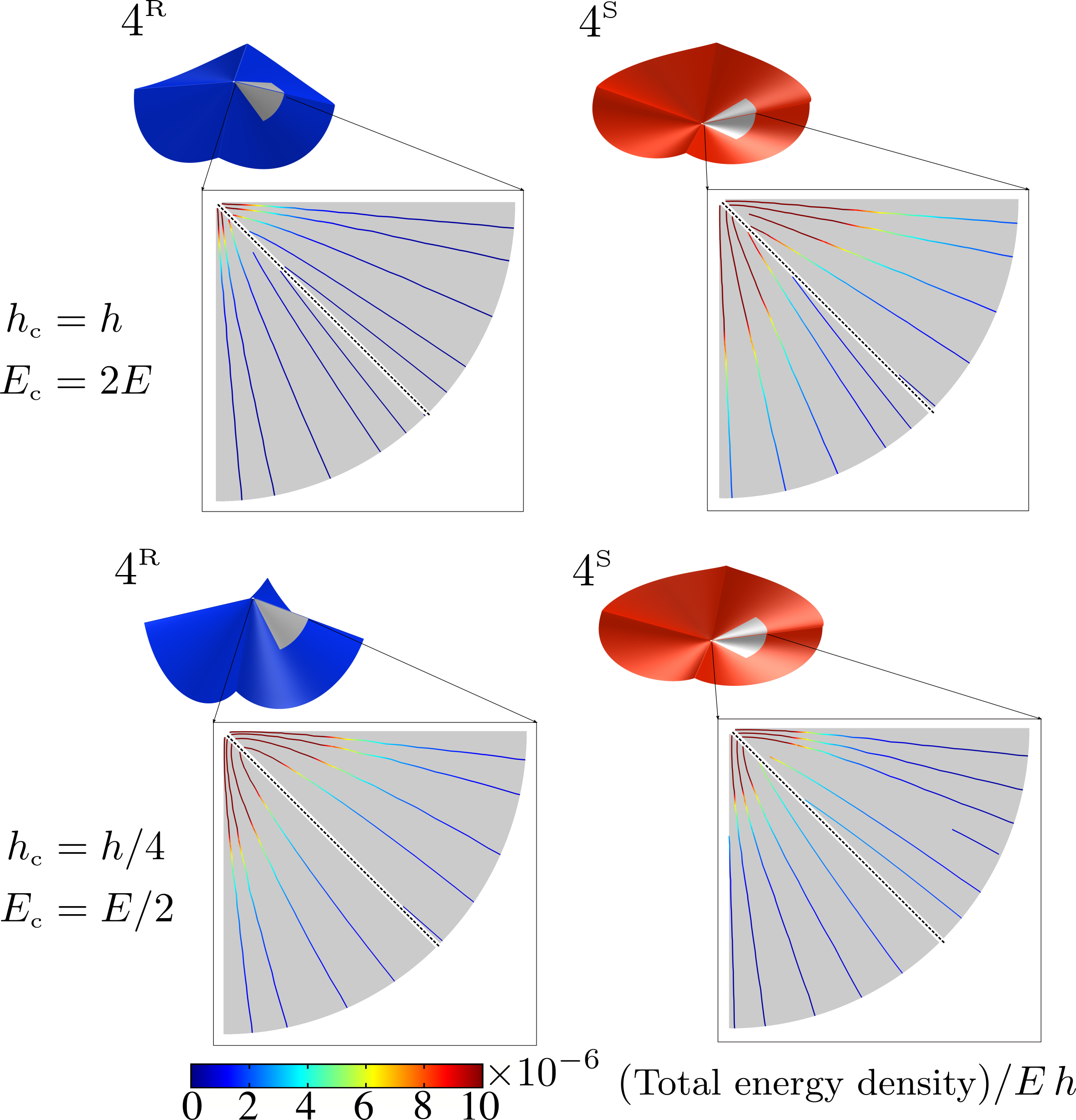}
\caption{Lines of smallest principal curvature of a symmetric foldable cone with four creases. The upper (resp. lower) row corresponds to a stiff (resp. soft) crease case. 
The equilibrium rest (left column) and snapped (right column) states are shown. 
The boxes show zoomed regions next to the vertex.
Black dotted line indicates the location of the crease and the color code corresponds to the elastic energy density.}
\label{Fig: Curvature lines}
\end{figure}

While the rest states are generally well predicted by the isometric constraint, the snapped states depart from the analytical predictions when $h_c$ is decreased.
This result could be explained in terms of the crease stiffness.
For a stiff crease, the folding angle is roughly constant along the crease enforcing the shape to be closer to a perfectly developable cone.
On the other hand, a $4^{\mbox{\tiny S}}$ cone is characterized by an azimuthal tension which favors stretching deformations of the panels and thus tensile traction on the crease. This mechanism could induce a varying folding angle along the crease and yield a structure that can depart from a perfect developable cone, especially for softer creases.
To verify this analysis, we plot in Fig.~\ref{Fig: Curvature lines} one quadrant containing the lines of smallest principal curvature for stiff and soft crease cases.
In a perfect developable cone, these lines coincide with the generators of a surface (\emph{i.e.} lines of zero curvature), however, we observe that the lines curve significantly close to the vertex, where the energy is concentrated.
This effect is more pronounced for the soft crease case than the stiff one. 

\begin{figure}[tbh]
\centering
\includegraphics[width=\columnwidth]{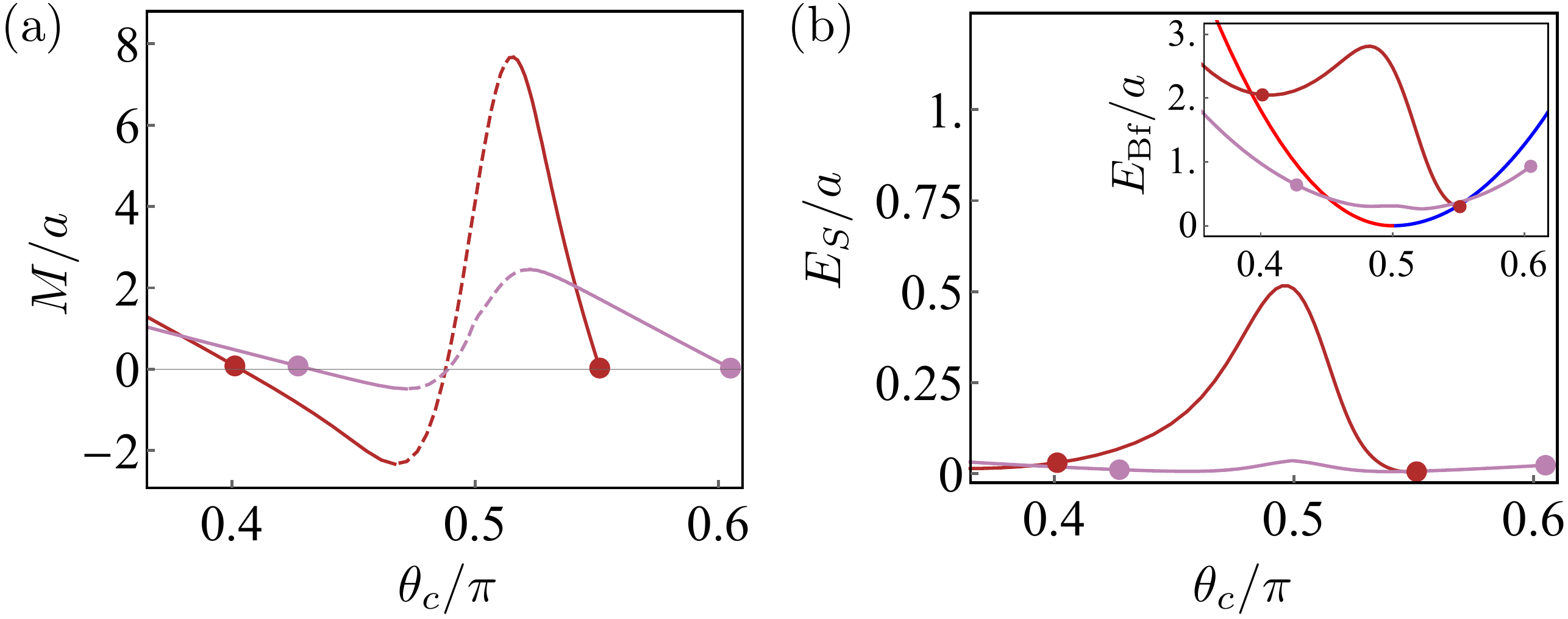}
\caption{(a) Moment $M$ as function of $\theta_c$ during the indentation of a stiff crease with $h_c=h$, $E_c=2E$ (\protect\colorfourline), and a soft crease, with $h_c=h/4$, $E_c=E/2$ (\protect\colortenline). (b) The corresponding total stretching energy. Inset. Normalized bending energy of the facets compared to the theoretical result of Fig.~\ref{Fig: Energy Landscape}(a).}
\label{Fig: NumericalBendingEnergy}
\end{figure}

Fig.~\ref{Fig: NumericalBendingEnergy}(a) shows typical curves for the moment as function of the polar angle $\theta_c$ during indentation for  both stiff and soft crease. 
Notice that the unstable branch of the stiffer crease is higher than that of the soft one, which in a real experiment leads to more energy being released during a snapping process. 
This observation can be attributed to a larger stretching energy barrier that is required to be overcome, which is evident from the energy plots shown in Fig.~\ref{Fig: NumericalBendingEnergy}(b).
To compare with analytical predictions, one should focus on the Inset of Fig.~\ref{Fig: NumericalBendingEnergy}(b) which shows the evolution of the bending energy of the facets throughout indentation and does not take into account the bending energy at the creases.
The predicted bending energy exhibits an asymmetric parabolic shape, where the minimum corresponds to a flat solution.
It is obvious that the bending energy of a soft crease follows closer the prediction than that of the stiff crease, which has a pronounced convex shape in the middle.

\section{Conclusion}

Foldable cones are the simplest example of a single-vertex origami whose facets can bend.
In the present work, we developed a theoretical model which allows us to obtain the shape of $f$-cones for any deflection.
The model shows that the bistable behavior of these structures is robust, regardless the specific properties of the creases.
In particular, for symmetrical all-mountain $f$-cones we obtained the polar angle at the crease as function of folding angle for both rest and snapped states.

However, in more realistic situations, the geometry and mechanical response of a $f$-cone are characterized by a competition between the elasticity of the facets (both their bending and stretching behavior) and the stiffness of the creases.
To this purpose, we have developed a continuous numerical model accounting for the both the elasticity of the creases and facets. Applying to the particular case of two perpendicular mountain creases we numerically studied the role of crease stiffness  and verified the snap-thorough behavior through a series of indentation tests. 
We studied the indentation paths in the $\theta_c(\psi)$ diagram and showed that the structures do not follow the shape of a perfect cone throughout the indentation.
For stiff creases, the path followed is that of an approximately constant folding angle while the two stable states lie closely to the theoretical prediction.
When the crease is made soft, the indentation paths follow closely the branches given by the isometrical constraints, however, it is noted that while the shape of the rest state is close to the theoretical prediction, the snapped one deviates further from it. 
From an energetic viewpoint, not only do stiffer creases lead to indentation paths with higher stretching energy barriers, but they also enforce the preferred angle more strongly. 
Hence, it can be concluded that a $f$-cone made with stiffer creases requires more stretching when passing through $\theta_c\sim\pi/2$. 
On the other hand, while softer creases induce large deviations of the preferred angles, they allow for low stretching during the inversion, which explains why they follow the bounds set by the analytical calculations more closely.

The present study validates our numerical model of temperature-induced hinge-like creases which can be applied to origami structures with more complex extended networks.
In this case, a temperature-induced folding of the network would work as a phase-field model where the sharp piecewise energy landscape is replaced by a smooth curve. The choice of temperature field as a trigger for crease formation is arbitrary as any other diffusion field, such as concentration~\cite{klein2007shaping} or swelling~\cite{kim2012designing}, would play a similar role. 
The main sought mechanism is to build up a reference configuration with a noneuclidean reference metric due to the presence of an initially imprinted crease network~\cite{wu2013three}. This approach is advantageous since one does not need to track sharp boundaries where the deformation fields are discontinuous. It renders numerical implementation tractable and less time-consuming, two important aspects when implementing the mechanical behavior of complex origami or crumpled structures.

\begin{acknowledgments}
I.A.-S. acknowledges the financial support of CONICYT DOCTORADO BECAS CHILE 2016-72170417. M.A.D. thanks the Velux Foundations for the support under the Villum Experiment program (project number 00023059). 
\end{acknowledgments}

\begin{widetext}
\begin{appendix}

\section{Derivation of the Euler Elastica Equation}
\label{Appendix: Functional variation}
The derivation of Euler's Elastica from the energy functional given by Eq.~(\ref{Functional panels}) can be found in~\cite{guven2008paper}. Is it instructive to repeat the calculations here for the sake of completeness.
The variation of the functional~\eqref{Functional panels} gives
\begin{eqnarray}
\delta F_{n} &=& \sum^{n}_{i=1} \int^{s_{i+1}}_{s_i} \left[  \left( -\kappa_i^{2} \mathbf{u}_i + \kappa_i \mathbf{n}_i + \lambda_i \mathbf{u}_i + \mathbf{f}'_i   \right) \cdot \delta \mathbf{u}_i +
\left( -(\kappa_i \mathbf{n}_i)' 
-  \kappa_i^2 \mathbf{t}_i+ \Lambda_i \mathbf{t}_i + \mathbf{f}_i \right) \cdot \delta \mathbf{t}_i \, ds \right]
 \nonumber\\
&&
+a \sum^{n}_{i=1}  (-\mathbf{f}_i \cdot \delta \mathbf{u}_i + \kappa_i \mathbf{n}_i \cdot \delta \mathbf{t}_i)\big|^{s_{i+1}}_{s_i} + \displaystyle \sum^{n}_{i=1} \delta g_i, \label{Functional variation 1}
\end{eqnarray}
where we have used the identities: $\mathbf{u}\cdot \mathbf{t} \times \delta \mathbf{t}' = \mathbf{n}\cdot \delta \mathbf{t}' $ and $\mathbf{u} \cdot \delta \mathbf{t} \times \mathbf{t}' = - \kappa \mathbf{t} \cdot \delta \mathbf{t}$.
The term $\delta g_i$ contributes to boundary terms only, and will be treated below.
Taking $\mathbf{u}_i$ and  $\mathbf{t}_i$ as independent variables, the terms proportional to $\delta \mathbf{u}_i$ yield
\begin{equation}
\mathbf{f}'_i = (\kappa_i^2 - \lambda_i) \mathbf{u}_i - a \kappa_i \mathbf{n}_i, \label{Eq: f' 1}
\end{equation}
while the terms proportional to $\delta \mathbf{t}_i$ give
\begin{equation}
\mathbf{f}_i= \kappa'\mathbf{n}_i + 2 \kappa_i^2 \mathbf{t}_i - \Lambda_i \mathbf{t}_i. \label{Eq: f 1}
\end{equation}
Notice that
\begin{eqnarray}
\mathbf{f}'_i \cdot \mathbf{t}_i &=& 0, \label{Ortogonal f derivado}\\
\mathbf{f}_i \cdot \mathbf{u}_i &=& 0 .\label{Ortogonal f}
\end{eqnarray}
Differentiating Eq.~\eqref{Eq: f 1} with respect to $s$ and using~\eqref{Ortogonal f derivado}, it follows that $\Lambda'_i = 5 \kappa_i \kappa'_i$. 
Integrating once, we obtain $\Lambda_i= 5\kappa_i^2/2 +c_i$, where $c_i$ is an integration constant.
Then, Eq.~\eqref{Eq: f 1} can be written as follows
\begin{equation}
\mathbf{f}_i = \kappa'_i \mathbf{n}_i - \left(\frac{1}{2}\kappa_i^2 + c_i \right)\mathbf{t}_i. \label{Appendix: f 2}
\end{equation}
Differentiating Eq.~(\ref{Appendix: f 2}) with respect to $s$ and projecting onto $\mathbf{u}_i$ gives
\begin{equation}
\mathbf{f}'_i \cdot \mathbf{u}_i = \frac{1}{2}\kappa_i^2 + c_i. \label{f projected onto u}
\end{equation}
Projecting equation~\eqref{Eq: f' 1} onto $\mathbf{u}_i$ and equating with Eq.~\eqref{f projected onto u}, one gets $\lambda_i = \kappa_i^2/2 - c_i$. Then, Eq.~\eqref{Eq: f' 1} now reads
\begin{equation}
\mathbf{f}'_i = \left(\frac{\kappa_i^2}{2} +c_i \right) \mathbf{u}_i -  \kappa_i \mathbf{n}_i. \label{f' 1}
\end{equation}
On the other hand, one can diefferentiate once Eq.~\eqref{Appendix: f 2} and obtain
\begin{equation}
\mathbf{f}'_i = \left[ \kappa''_i +\kappa_i \left(\frac{\kappa_i^2}{2} +c_i \right) \right]  \mathbf{n}_i + \left(\frac{\kappa_i^2}{2} +c_i \right) \mathbf{u}_i. \label{f' 2}
\end{equation}
Using equations~(\ref{f' 1},\ref{f' 2}), one obtains the Euler's \emph{Elastica} equations given by Eq.~\eqref{Equation for kappa i}.
In the following, the boundary terms will be treated.

\section{Boundary Conditions}
\label{Appendix: Boundary Conditions}

It is useful to compute the variation of the functional \eqref{Functional panels} in terms of virtual rotations of the frame specified by the Euler-like angles. To this purpose,
we first introduce the vectors $\mathbf{e}_{\varphi} =  -\sin{\varphi}\, \mathbf{x} + \cos{\varphi}\, \mathbf{y} $ and $\mathbf{n}_\varphi \equiv \mathbf{u} \times \mathbf{e}_\varphi$ which span the plane containing the vectors $\mathbf{t}$ and $\mathbf{n}$ (for simplicity, we omit subscripts here). 
If $\phi(s)$ is the angle between $\mathbf{t}$ and $\mathbf{e}_\varphi$, then,
\begin{subequations}
 \begin{eqnarray}
\mathbf{t} &=& \cos  \phi \, \mathbf{e}_\varphi + \sin  \phi \, \mathbf{n}_\varphi, \\
\mathbf{n} &=& -\sin  \phi \, \mathbf{e}_\varphi + \cos  \phi \, \mathbf{n}_\varphi.
\end{eqnarray}
\end{subequations}
Defining $\mathbf{e}_\rho = \cos{\varphi}\, \mathbf{x} + \sin{\varphi}\, \mathbf{y}$, one can write
\begin{subequations}
 \begin{eqnarray}
\mathbf{u} &=& \sin \theta \, \mathbf{e}_\rho + \cos{\theta} \, \mathbf{z}, \\
\mathbf{n}_\varphi &=& - \cos{\theta} \, \mathbf{e}_\rho + \sin \theta \, \mathbf{z}.
\end{eqnarray}
\end{subequations}
Then, one can show the following relations
\begin{eqnarray}
\delta \mathbf{u} &=& - \delta \theta \,\mathbf{n}_\varphi + \sin \theta \,\delta \varphi \,\mathbf{e}_\varphi, \\
\delta \mathbf{t} &=&  ( \delta \phi + \cos{\theta} \, \delta \varphi) \mathbf{n} + (\sin{\phi} \, \delta \theta - \sin{\theta} \cos{\phi} \, \delta \varphi) \mathbf{u}.
\end{eqnarray}
Notice that $\mathbf{u}\cdot \delta \mathbf{u}=0$ and $\mathbf{t}\cdot \delta \mathbf{t}=0$ as expected. The following relations are useful
\begin{eqnarray}
\mathbf{n}\cdot \delta \mathbf{u} &=&  -\cos{\phi} \, \delta \theta - \sin{\theta} \sin{\phi} \, \delta \varphi  ,\nonumber \\
\mathbf{t}\cdot \delta \mathbf{u} &=&  -\sin{\phi} \, \delta \theta + \sin{\theta} \cos{\phi} \, \delta \varphi ,\nonumber \\
\mathbf{n}\cdot \delta \mathbf{t} &=& \delta \phi + \cos{\theta} \, \delta \varphi .\label{Explicit variations}
\end{eqnarray}

Now, we put the subscripts back and write some useful relations.
First, notice that in the plane perpendicular to a crease, the frame $\{ \mathbf{t},\mathbf{n} \}$ rotates by an angle $\psi_i-\pi$, which can be expressed as follows
\begin{equation}
\begin{pmatrix}
\mathbf{t}^{+}_i  \\
\mathbf{n}^{+}_i 
\end{pmatrix}
=\begin{pmatrix}
- \cos{\psi_i} &  \sin{\psi_i}  \\
 -\sin{\psi_i} & - \cos{\psi_i} 
\end{pmatrix}
\begin{pmatrix}
\mathbf{t}^{-}_i  \\
\mathbf{n}^{-}_i 
\end{pmatrix}. \label{Frame Rotation}
\end{equation}

Using the relations given by Eq.~\eqref{Explicit variations} and Eq.~\eqref{Frame Rotation} one obtains the following relations
\begin{eqnarray}
\mathbf{n}^{+}_i \cdot \delta \mathbf{t}^{+}_i  &=&  \delta \phi^{+}_i  + \cos{\theta^{+}_i} \, \delta \varphi^{+}_i ,  
\nonumber\\
\mathbf{n}^{-}_i \cdot \delta \mathbf{t}^{-}_i  &=& \delta  \phi^{-}_i  + \cos{\theta^{+}_i} \, \delta \varphi^{+}_i , 
\nonumber\\
\mathbf{t}^{+}_i  \cdot \delta  \mathbf{t}^{-}_i  &=& - \sin{\psi_i} \left(\delta  \phi^{-}_i  + \cos{\theta^{+}_i} \, \delta \varphi^{+}_i \right),  \nonumber\\
 \mathbf{t}^{-}_i \cdot \delta \mathbf{t}^{+}_i   &=&  \sin{\psi_i}  \left(\delta  \phi^{+}_i  + \cos{\theta^{+}_i} \, \delta \varphi^{+}_i \right),  \nonumber\\
\mathbf{n}^{+}_i \cdot \delta \mathbf{t}^{-}_i  &=& -\cos{\psi_i} \left( \delta \phi^{-}_i  + \cos{\theta^{+}_i} \, \delta \varphi^{+}_i \right),  \nonumber\\
\mathbf{n}^{-}_i \cdot \delta \mathbf{t}^{+}_i  &=& -\cos{\psi_i} \left(\delta  \phi^{+}_i  + \cos{\theta^{+}_i} \, \delta \varphi^{+}_i \right).
\label{eq:minus2}
\end{eqnarray}
where $\phi^{\pm}_i$ is the angle between $\mathbf{t}^{\pm}_i $ and $\mathbf{e}_{\theta}$. 
Notice that $\psi_i = \pi + \phi^{+}_i - \phi^{-}_i$, then, $\delta \psi_i = \delta  \phi^{+}_i  - \delta  \phi^{-}_i $. 
At this stage, we distinguish two cases: infinitely stiff creases and finite crease stiffness.

\subsection{Infinitely stiff crease}

By taking $\delta g_i=0$ and using the periodic convention in Eq.~\eqref{Eq: Boundary terms + Crease energy}, one can write
\begin{equation}
\sum^{n}_{i=1} \left[ \left( \mathbf{f}^{+}_i -  \mathbf{f}^{-}_i \right)\cdot \delta \mathbf{u}^{+}_i +   \kappa^{-}_i  \mathbf{n}^{-}_i \cdot  \delta \mathbf{t}^{-}_i -  \kappa^{+}_i \mathbf{n}^{+}_i \cdot \delta \mathbf{t}^{+}_i \right] = 0,
\label{Eq: Boundary terms after periodic convention}
\end{equation}
where we have used $\mathbf{u}^{-}_i=~\mathbf{u}^{+}_i$.
Using Eqs.~(\ref{eq:minus2}), the condition of infinitely stiff crease $\delta \psi_i = 0$ is equivalent to imposing $\mathbf{n}^{+}_i \cdot \delta \mathbf{t}^{-}_i =\mathbf{n}^{-}_i \cdot \delta \mathbf{t}^{+}_i  $.
Thus, imposing $\delta \psi_i=0$ and letting $\delta \mathbf{u}^{+}_i$ undergo independent virtual rotation imply the boundary conditions~\eqref{Eq: Curvature continuity} and~\eqref{Eq: Continuity f}.
By projecting Eq.~\eqref{Eq: Continuity f} onto $\mathbf{n}^{+}_i$ and $\mathbf{t}^{+}_i$, one obtains
\begin{equation}
\begin{pmatrix}
\kappa^{\prime +}_i \\
{\kappa^{+2}_i}/2+ c 
\end{pmatrix}
=
\begin{pmatrix}
- \cos{\psi_i} &  -\sin{\psi_i} \\
 \sin{\psi_i} & -\cos{\psi_i}
\end{pmatrix}
\begin{pmatrix}
\kappa^{\prime -}_i \\
\kappa^{-2}_i/2+ c
\end{pmatrix}, \label{Eq: Frame Rotation}
\end{equation}
where we have assumed $c_i=c$.
Manipulating Eq.~\eqref{Eq: Frame Rotation} one obtains equations~\eqref{Eq: Condition prime} and~\eqref{Eq: Condition psi}.

\subsection{Finite crease stiffness}

The variation of the crease energy given by Eq.~\eqref{Crease energy} reads
\begin{equation}
\delta g_i = L \, \left[ \sigma_i \left( \delta  \mathbf{t}^{-}_i \cdot \mathbf{t}^{+}_i  +  \mathbf{t}^{-}_i \cdot \delta \mathbf{t}^{+}_i  \right) + \tau_i \left( \left(\delta \mathbf{t}^{-}_i \times \mathbf{t}^{+}_i  \right) \cdot \mathbf{u}^{+}_i+  \left( \mathbf{t}^{-}_i \times \delta \mathbf{t}^{+}_i  \right) \cdot \mathbf{u}^{+}_i +   \left( \mathbf{t}^{-}_i \times \mathbf{t}^{+}_i  \right) \cdot \delta \mathbf{u}^{+}_i\right) \right] .
\end{equation}
The last term in the right-hand side is zero because $( \mathbf{t}^{-}_i \times \mathbf{t}^{+}_i  ) |_{s_i}$ is parallel to $\mathbf{u}_i (s_i)$. 
Using the cyclic properties of the triple product we can write
\begin{equation}
\delta g_i = L \, \left[ \sigma_i \left( \mathbf{t}^{+}_i  \cdot \delta  \mathbf{t}^{-}_i+  \mathbf{t}^{-}_i \cdot \delta \mathbf{t}^{+}_i  \right) + \tau_i \left( \mathbf{n}^{-}_i \cdot \delta \mathbf{t}^{+}_i  - \mathbf{n}^{+}_i \cdot \delta \mathbf{t}^{-}_i \right) \right] . \label{App: Crease variation}
\end{equation}
Using the identities~(\ref{eq:minus2}),  Eq.~\eqref{App: Crease variation} can be rewritten as follows
\begin{equation}
\delta g_i = L \left[\sigma_i \, \sin{\psi_i}\left(\delta  \phi^{+}_i  - \delta  \phi^{-}_i  \right) + \tau_i \, \cos{\psi_i}\left( \delta  \phi^{+}_i  - \delta  \phi^{-}_i  \right) \right].
\end{equation}
Notice that the above equation has the form $\delta g_i = (d g_i / d \psi_i)\delta \psi_i$.

Using the definition for the constants introduced in Eq.~\eqref{Crease constants} and recalling that $\theta^{+}_i=\theta^{-}_i$ and $\varphi^{+}_i=\varphi^{-}_i$, we can rewrite Eq.~\eqref{Eq: Boundary terms + Crease energy} as follows
\begin{eqnarray}
&&\sum^{n}_{i=1} \left\lbrace \left( -\kappa^{\prime +}_i  \cos{\phi^{+}_i} +\kappa^{\prime-}_i  \cos{\phi^{-}_i}  
+  \left(\frac{1}{2}{\kappa^{+}_i}^{2} + c \right)\sin{\phi^{+}_i} -  \left(\frac{1}{2}{\kappa^{-}_i}^{2} + c \right)\sin{\phi^{-}_i} \right) \delta \theta^{+}_i \right.  \nonumber \\
&& + \left( -\kappa^{\prime +}_i  \sin{\phi^{+}_i} + \kappa^{\prime-}_i  \sin{\phi^{-}_i} 
-  \left(\frac{1}{2}{\kappa^{+}_i}^{2} + c \right)\cos{\phi^{+}_i} +  \left(\frac{1}{2}{\kappa^{-}_i}^{2} + c \right)\cos{\phi^{-}_i} \right)\sin{\theta^{+}_i} \delta \varphi^{+}_i 
- \left( \kappa^{+}_i - \kappa^{-}_i\right)  \cos{\theta^{+}_i} \delta \varphi^{+}_i 
\nonumber \\
&& \left. - \left( \kappa^{+}_i- \bar{k}_i \sin \left( \psi_i - \psi_i^{0} \right)\right) \delta \phi^{+}_i  + \left( \kappa^{-}_i- \bar{k}_i \sin \left( \psi_i - \psi_i^{0} \right)\right) \delta \phi^{-}_i\right\rbrace =0.
\end{eqnarray}
The infinitesimal variations of the frame vectors can be translated to virtual rotations in terms of the Euler angles $\delta \theta^{+}_i$, $\delta \varphi^{+}_i$, $\delta \phi^{+}_i$ and $\delta \phi^{-}_i$. 
Assuming that all these virtual rotations are independent, one obtains conditions \eqref{Eq: Curvature continuity}, \eqref{Eq: Continuity f} and \eqref{Crease energy boundary condition} with $g_i (\psi_i)$ given by Eq.~(\ref{Crease energy}).

\section{Closure condition}
\label{Appendix: Closure condition}
In a symmetrical $f$-cone with $n$ creases, the  azimuthal angle spanned by a single panel is given by the integral
\begin{equation}
\Delta \varphi =\int^{\tfrac{\alpha}{2}}_{-\tfrac{\alpha}{2}}  \frac{\kappa^2/2 + c}{J\sin^2 \theta} ds =\int^{\tfrac{\alpha}{2}}_{-\tfrac{\alpha}{2}}  \frac{J}{2} \left[ \frac{J^{2} + 2c}{J^{2} - \kappa^{2}} -1\right] \, ds,\label{Closure condition 2}
\end{equation}
where $\alpha=2 \pi /n$. 
As the deformed state will also be symmetrical, the closure condition can be written as $\Delta \varphi = \pm \alpha$ or $\Delta \varphi =0$ according to Eq.~(\ref{eq:closure}).
These integrals can be computed analytically for each state:
\begin{equation}
\Delta \varphi^{\mbox{\tiny R}} = 
\left[ \frac{J(J^{2} + 2 c) }{ \kappa_{0} (J^{2} - \kappa^{2}_{0} ) } \sqrt{m_{r}} \, \Pi \left( \frac{\kappa^{2}_{0} }{ \kappa^{2}_{0} -J^{2}}  ,\text{am}\left( \frac{\kappa_{0}}{2 \sqrt{m_{r}}} s \Big| m_{r} \right) \Big| m_{r} \right)-  \frac{J}{2}s \right]\Bigg|^{\tfrac{\alpha}{2}}_{-\tfrac{\alpha}{2}}, 
\label{Eq: Integral closure condition rest}
\end{equation}
and
\begin{equation}
\Delta \varphi^{\mbox{\tiny S}}= \left[ \frac{c }{J}s -\frac{(J^{2} + 2 c )\kappa_{0}(m_{s}-1)}{ J(J^{2} +(m_{s}-1) \kappa^{2}_{0} ) } \Pi \left( \frac{ J^{2} m_{s} }{J^{2} +(m_{s}-1) \kappa^{2}_{0} } ,\text{am}\left( \frac{\kappa_{0}}{2}s \Big| m_{s} \right) \Bigg| m_{s} \right) \right]\Bigg|^{\tfrac{\alpha}{2}}_{-\tfrac{\alpha}{2}},
\label{Eq: Integral closure condition snap}
\end{equation}
where the labels $\mbox{R},\mbox{S}$ stand, respectively, for the rest and snapped states. Also, $\Pi (\cdot,\cdot | m)$ is the elliptic integral of third kind and $\text{am}(\cdot | m)$ is the Jacobi amplitude with modulus $m$~\cite{abramowitz1965handbook}. 

\end{appendix}
\end{widetext}

\bibliography{GeneralModelFCone}

\begin{thebibliography}{32}%
\makeatletter
\providecommand \@ifxundefined [1]{%
 \@ifx{#1\undefined}
}%
\providecommand \@ifnum [1]{%
 \ifnum #1\expandafter \@firstoftwo
 \else \expandafter \@secondoftwo
 \fi
}%
\providecommand \@ifx [1]{%
 \ifx #1\expandafter \@firstoftwo
 \else \expandafter \@secondoftwo
 \fi
}%
\providecommand \natexlab [1]{#1}%
\providecommand \enquote  [1]{``#1''}%
\providecommand \bibnamefont  [1]{#1}%
\providecommand \bibfnamefont [1]{#1}%
\providecommand \citenamefont [1]{#1}%
\providecommand \href@noop [0]{\@secondoftwo}%
\providecommand \href [0]{\begingroup \@sanitize@url \@href}%
\providecommand \@href[1]{\@@startlink{#1}\@@href}%
\providecommand \@@href[1]{\endgroup#1\@@endlink}%
\providecommand \@sanitize@url [0]{\catcode `\\12\catcode `\$12\catcode
  `\&12\catcode `\#12\catcode `\^12\catcode `\_12\catcode `\%12\relax}%
\providecommand \@@startlink[1]{}%
\providecommand \@@endlink[0]{}%
\providecommand \url  [0]{\begingroup\@sanitize@url \@url }%
\providecommand \@url [1]{\endgroup\@href {#1}{\urlprefix }}%
\providecommand \urlprefix  [0]{URL }%
\providecommand \Eprint [0]{\href }%
\providecommand \doibase [0]{http://dx.doi.org/}%
\providecommand \selectlanguage [0]{\@gobble}%
\providecommand \bibinfo  [0]{\@secondoftwo}%
\providecommand \bibfield  [0]{\@secondoftwo}%
\providecommand \translation [1]{[#1]}%
\providecommand \BibitemOpen [0]{}%
\providecommand \bibitemStop [0]{}%
\providecommand \bibitemNoStop [0]{.\EOS\space}%
\providecommand \EOS [0]{\spacefactor3000\relax}%
\providecommand \BibitemShut  [1]{\csname bibitem#1\endcsname}%
\let\auto@bib@innerbib\@empty
%</preamble>
\bibitem [{\citenamefont {Schenk}\ and\ \citenamefont
  {Guest}(2013)}]{Schenk3276}%
  \BibitemOpen
  \bibfield  {author} {\bibinfo {author} {\bibfnamefont {M.}~\bibnamefont
  {Schenk}}\ and\ \bibinfo {author} {\bibfnamefont {S.~D.}\ \bibnamefont
  {Guest}},\ }\href {\doibase 10.1073/pnas.1217998110} {\bibfield  {journal}
  {\bibinfo  {journal} {Proceedings of the National Academy of Sciences}\
  }\textbf {\bibinfo {volume} {110}},\ \bibinfo {pages} {3276} (\bibinfo {year}
  {2013})}\BibitemShut {NoStop}%
\bibitem [{\citenamefont {Wei}\ \emph {et~al.}(2013)\citenamefont {Wei},
  \citenamefont {Guo}, \citenamefont {Dudte}, \citenamefont {Liang},\ and\
  \citenamefont {Mahadevan}}]{wei2013geometric}%
  \BibitemOpen
  \bibfield  {author} {\bibinfo {author} {\bibfnamefont {Z.~Y.}\ \bibnamefont
  {Wei}}, \bibinfo {author} {\bibfnamefont {Z.~V.}\ \bibnamefont {Guo}},
  \bibinfo {author} {\bibfnamefont {L.}~\bibnamefont {Dudte}}, \bibinfo
  {author} {\bibfnamefont {H.~Y.}\ \bibnamefont {Liang}}, \ and\ \bibinfo
  {author} {\bibfnamefont {L.}~\bibnamefont {Mahadevan}},\ }\href@noop {}
  {\bibfield  {journal} {\bibinfo  {journal} {Physical Review Letters}\
  }\textbf {\bibinfo {volume} {110}},\ \bibinfo {pages} {215501} (\bibinfo
  {year} {2013})}\BibitemShut {NoStop}%
\bibitem [{\citenamefont {Miura}(1985)}]{miura1985method}%
  \BibitemOpen
  \bibfield  {author} {\bibinfo {author} {\bibfnamefont {K.}~\bibnamefont
  {Miura}},\ }\href@noop {} {\emph {\bibinfo {title} {Method of Packaging and
  Deployment of Large Membranes in Space}}},\ \bibinfo {type} {Tech. Rep.}\
  \bibinfo {number} {618}\ (\bibinfo  {institution} {The Institute of Space and
  Astronautical Science},\ \bibinfo {year} {1985})\BibitemShut {NoStop}%
\bibitem [{\citenamefont {Dias}\ \emph {et~al.}(2012)\citenamefont {Dias},
  \citenamefont {Dudte}, \citenamefont {Mahadevan},\ and\ \citenamefont
  {Santangelo}}]{dias2012geometric}%
  \BibitemOpen
  \bibfield  {author} {\bibinfo {author} {\bibfnamefont {M.~A.}\ \bibnamefont
  {Dias}}, \bibinfo {author} {\bibfnamefont {L.~H.}\ \bibnamefont {Dudte}},
  \bibinfo {author} {\bibfnamefont {L.}~\bibnamefont {Mahadevan}}, \ and\
  \bibinfo {author} {\bibfnamefont {C.~D.}\ \bibnamefont {Santangelo}},\
  }\href@noop {} {\bibfield  {journal} {\bibinfo  {journal} {Physical Review
  Letters}\ }\textbf {\bibinfo {volume} {109}},\ \bibinfo {pages} {114301}
  (\bibinfo {year} {2012})}\BibitemShut {NoStop}%
\bibitem [{\citenamefont {Dudte}\ \emph {et~al.}(2016)\citenamefont {Dudte},
  \citenamefont {Vouga}, \citenamefont {Tachi},\ and\ \citenamefont
  {Mahadevan}}]{dudte2016programming}%
  \BibitemOpen
  \bibfield  {author} {\bibinfo {author} {\bibfnamefont {L.~H.}\ \bibnamefont
  {Dudte}}, \bibinfo {author} {\bibfnamefont {E.}~\bibnamefont {Vouga}},
  \bibinfo {author} {\bibfnamefont {T.}~\bibnamefont {Tachi}}, \ and\ \bibinfo
  {author} {\bibfnamefont {L.}~\bibnamefont {Mahadevan}},\ }\href@noop {}
  {\bibfield  {journal} {\bibinfo  {journal} {Nature Materials}\ }\textbf
  {\bibinfo {volume} {15}},\ \bibinfo {pages} {583} (\bibinfo {year}
  {2016})}\BibitemShut {NoStop}%
\bibitem [{\citenamefont {Silverberg}\ \emph {et~al.}(2014)\citenamefont
  {Silverberg}, \citenamefont {Evans}, \citenamefont {McLeod}, \citenamefont
  {Hayward}, \citenamefont {Hull}, \citenamefont {Santangelo},\ and\
  \citenamefont {Cohen}}]{silverberg2014using}%
  \BibitemOpen
  \bibfield  {author} {\bibinfo {author} {\bibfnamefont {J.~L.}\ \bibnamefont
  {Silverberg}}, \bibinfo {author} {\bibfnamefont {A.~A.}\ \bibnamefont
  {Evans}}, \bibinfo {author} {\bibfnamefont {L.}~\bibnamefont {McLeod}},
  \bibinfo {author} {\bibfnamefont {R.~C.}\ \bibnamefont {Hayward}}, \bibinfo
  {author} {\bibfnamefont {T.}~\bibnamefont {Hull}}, \bibinfo {author}
  {\bibfnamefont {C.~D.}\ \bibnamefont {Santangelo}}, \ and\ \bibinfo {author}
  {\bibfnamefont {I.}~\bibnamefont {Cohen}},\ }\href@noop {} {\bibfield
  {journal} {\bibinfo  {journal} {Science}\ }\textbf {\bibinfo {volume}
  {345}},\ \bibinfo {pages} {647} (\bibinfo {year} {2014})}\BibitemShut
  {NoStop}%
\bibitem [{\citenamefont {Waitukaitis}\ \emph {et~al.}(2015)\citenamefont
  {Waitukaitis}, \citenamefont {Menaut}, \citenamefont {Chen},\ and\
  \citenamefont {van Hecke}}]{waitukaitis2015origami}%
  \BibitemOpen
  \bibfield  {author} {\bibinfo {author} {\bibfnamefont {S.}~\bibnamefont
  {Waitukaitis}}, \bibinfo {author} {\bibfnamefont {R.}~\bibnamefont {Menaut}},
  \bibinfo {author} {\bibfnamefont {B.~G.}\ \bibnamefont {Chen}}, \ and\
  \bibinfo {author} {\bibfnamefont {M.}~\bibnamefont {van Hecke}},\ }\href@noop
  {} {\bibfield  {journal} {\bibinfo  {journal} {Physical Review Letters}\
  }\textbf {\bibinfo {volume} {114}},\ \bibinfo {pages} {055503} (\bibinfo
  {year} {2015})}\BibitemShut {NoStop}%
\bibitem [{\citenamefont {Boatti}\ \emph {et~al.}(2017)\citenamefont {Boatti},
  \citenamefont {Vasios},\ and\ \citenamefont {Bertoldi}}]{boatti2017origami}%
  \BibitemOpen
  \bibfield  {author} {\bibinfo {author} {\bibfnamefont {E.}~\bibnamefont
  {Boatti}}, \bibinfo {author} {\bibfnamefont {N.}~\bibnamefont {Vasios}}, \
  and\ \bibinfo {author} {\bibfnamefont {K.}~\bibnamefont {Bertoldi}},\
  }\href@noop {} {\bibfield  {journal} {\bibinfo  {journal} {Advanced
  Materials}\ }\textbf {\bibinfo {volume} {29}},\ \bibinfo {pages} {1700360}
  (\bibinfo {year} {2017})}\BibitemShut {NoStop}%
\bibitem [{\citenamefont {Pratapa}\ \emph {et~al.}(2019)\citenamefont
  {Pratapa}, \citenamefont {Liu},\ and\ \citenamefont
  {Paulino}}]{PhysRevLett.122.155501}%
  \BibitemOpen
  \bibfield  {author} {\bibinfo {author} {\bibfnamefont {P.~P.}\ \bibnamefont
  {Pratapa}}, \bibinfo {author} {\bibfnamefont {K.}~\bibnamefont {Liu}}, \ and\
  \bibinfo {author} {\bibfnamefont {G.~H.}\ \bibnamefont {Paulino}},\ }\href
  {\doibase 10.1103/PhysRevLett.122.155501} {\bibfield  {journal} {\bibinfo
  {journal} {Physical Review Letters}\ }\textbf {\bibinfo {volume} {122}},\
  \bibinfo {pages} {155501} (\bibinfo {year} {2019})}\BibitemShut {NoStop}%
\bibitem [{\citenamefont {Hanna}\ \emph {et~al.}(2014)\citenamefont {Hanna},
  \citenamefont {Lund}, \citenamefont {Lang}, \citenamefont {Magleby},\ and\
  \citenamefont {Howell}}]{hanna2014waterbomb}%
  \BibitemOpen
  \bibfield  {author} {\bibinfo {author} {\bibfnamefont {B.~H.}\ \bibnamefont
  {Hanna}}, \bibinfo {author} {\bibfnamefont {J.~M.}\ \bibnamefont {Lund}},
  \bibinfo {author} {\bibfnamefont {R.~J.}\ \bibnamefont {Lang}}, \bibinfo
  {author} {\bibfnamefont {S.~P.}\ \bibnamefont {Magleby}}, \ and\ \bibinfo
  {author} {\bibfnamefont {L.~L.}\ \bibnamefont {Howell}},\ }\href@noop {}
  {\bibfield  {journal} {\bibinfo  {journal} {Smart Materials and Structures}\
  }\textbf {\bibinfo {volume} {23}},\ \bibinfo {pages} {094009} (\bibinfo
  {year} {2014})}\BibitemShut {NoStop}%
\bibitem [{\citenamefont {Brunck}\ \emph {et~al.}(2016)\citenamefont {Brunck},
  \citenamefont {Lechenault}, \citenamefont {Reid},\ and\ \citenamefont
  {Adda-Bedia}}]{brunck2016elastic}%
  \BibitemOpen
  \bibfield  {author} {\bibinfo {author} {\bibfnamefont {V.}~\bibnamefont
  {Brunck}}, \bibinfo {author} {\bibfnamefont {F.}~\bibnamefont {Lechenault}},
  \bibinfo {author} {\bibfnamefont {A.}~\bibnamefont {Reid}}, \ and\ \bibinfo
  {author} {\bibfnamefont {M.}~\bibnamefont {Adda-Bedia}},\ }\href@noop {}
  {\bibfield  {journal} {\bibinfo  {journal} {Physical Review E}\ }\textbf
  {\bibinfo {volume} {93}},\ \bibinfo {pages} {033005} (\bibinfo {year}
  {2016})}\BibitemShut {NoStop}%
\bibitem [{\citenamefont {Chen}\ \emph {et~al.}(2015)\citenamefont {Chen},
  \citenamefont {Peng},\ and\ \citenamefont {You}}]{chen2015origami}%
  \BibitemOpen
  \bibfield  {author} {\bibinfo {author} {\bibfnamefont {Y.}~\bibnamefont
  {Chen}}, \bibinfo {author} {\bibfnamefont {R.}~\bibnamefont {Peng}}, \ and\
  \bibinfo {author} {\bibfnamefont {Z.}~\bibnamefont {You}},\ }\href@noop {}
  {\bibfield  {journal} {\bibinfo  {journal} {Science}\ }\textbf {\bibinfo
  {volume} {349}},\ \bibinfo {pages} {396} (\bibinfo {year}
  {2015})}\BibitemShut {NoStop}%
\bibitem [{\citenamefont {Lechenault}\ \emph {et~al.}(2014)\citenamefont
  {Lechenault}, \citenamefont {Thiria},\ and\ \citenamefont
  {Adda-Bedia}}]{lechenault2014mechanical}%
  \BibitemOpen
  \bibfield  {author} {\bibinfo {author} {\bibfnamefont {F.}~\bibnamefont
  {Lechenault}}, \bibinfo {author} {\bibfnamefont {B.}~\bibnamefont {Thiria}},
  \ and\ \bibinfo {author} {\bibfnamefont {M.}~\bibnamefont {Adda-Bedia}},\
  }\href@noop {} {\bibfield  {journal} {\bibinfo  {journal} {Physical Review
  Letters}\ }\textbf {\bibinfo {volume} {112}},\ \bibinfo {pages} {244301}
  (\bibinfo {year} {2014})}\BibitemShut {NoStop}%
\bibitem [{\citenamefont {Lechenault}\ and\ \citenamefont
  {Adda-Bedia}(2015)}]{lechenault2015generic}%
  \BibitemOpen
  \bibfield  {author} {\bibinfo {author} {\bibfnamefont {F.}~\bibnamefont
  {Lechenault}}\ and\ \bibinfo {author} {\bibfnamefont {M.}~\bibnamefont
  {Adda-Bedia}},\ }\href@noop {} {\bibfield  {journal} {\bibinfo  {journal}
  {Physical Review Letters}\ }\textbf {\bibinfo {volume} {115}},\ \bibinfo
  {pages} {235501} (\bibinfo {year} {2015})}\BibitemShut {NoStop}%
\bibitem [{\citenamefont {Huffman}(1976)}]{huffman1976curvature}%
  \BibitemOpen
  \bibfield  {author} {\bibinfo {author} {\bibfnamefont {D.~A.}\ \bibnamefont
  {Huffman}},\ }\href@noop {} {\bibfield  {journal} {\bibinfo  {journal} {IEEE
  Transactions on Computers}\ }\textbf {\bibinfo {volume} {C-25}},\ \bibinfo
  {pages} {1010} (\bibinfo {year} {1976})}\BibitemShut {NoStop}%
\bibitem [{\citenamefont {Seffen}(2016)}]{seffen2016fundamental}%
  \BibitemOpen
  \bibfield  {author} {\bibinfo {author} {\bibfnamefont {K.~A.}\ \bibnamefont
  {Seffen}},\ }\href@noop {} {\bibfield  {journal} {\bibinfo  {journal}
  {Physical Review E}\ }\textbf {\bibinfo {volume} {94}},\ \bibinfo {pages}
  {013002} (\bibinfo {year} {2016})}\BibitemShut {NoStop}%
\bibitem [{\citenamefont {Guven}\ \emph {et~al.}(2011)\citenamefont {Guven},
  \citenamefont {M{\"u}ller},\ and\ \citenamefont
  {V{\'a}zquez-Montejo}}]{guven2011conical}%
  \BibitemOpen
  \bibfield  {author} {\bibinfo {author} {\bibfnamefont {J.}~\bibnamefont
  {Guven}}, \bibinfo {author} {\bibfnamefont {M.~M.}\ \bibnamefont
  {M{\"u}ller}}, \ and\ \bibinfo {author} {\bibfnamefont {P.}~\bibnamefont
  {V{\'a}zquez-Montejo}},\ }\href@noop {} {\bibfield  {journal} {\bibinfo
  {journal} {Journal of Physics A: Mathematical and Theoretical}\ }\textbf
  {\bibinfo {volume} {45}},\ \bibinfo {pages} {015203} (\bibinfo {year}
  {2011})}\BibitemShut {NoStop}%
\bibitem [{\citenamefont {Ben~Amar}\ and\ \citenamefont
  {Pomeau}(1997)}]{ben1997crumpled}%
  \BibitemOpen
  \bibfield  {author} {\bibinfo {author} {\bibfnamefont {M.}~\bibnamefont
  {Ben~Amar}}\ and\ \bibinfo {author} {\bibfnamefont {Y.}~\bibnamefont
  {Pomeau}},\ }\href@noop {} {\bibfield  {journal} {\bibinfo  {journal}
  {Proceedings of the Royal Society of London. Series A: Mathematical, Physical
  and Engineering Sciences}\ }\textbf {\bibinfo {volume} {453}},\ \bibinfo
  {pages} {729} (\bibinfo {year} {1997})}\BibitemShut {NoStop}%
\bibitem [{\citenamefont {Cerda}\ and\ \citenamefont
  {Mahadevan}(1998)}]{cerda1998conical}%
  \BibitemOpen
  \bibfield  {author} {\bibinfo {author} {\bibfnamefont {E.}~\bibnamefont
  {Cerda}}\ and\ \bibinfo {author} {\bibfnamefont {L.}~\bibnamefont
  {Mahadevan}},\ }\href@noop {} {\bibfield  {journal} {\bibinfo  {journal}
  {Physical Review Letters}\ }\textbf {\bibinfo {volume} {80}},\ \bibinfo
  {pages} {2358} (\bibinfo {year} {1998})}\BibitemShut {NoStop}%
\bibitem [{\citenamefont {Cerda}\ \emph {et~al.}(1999)\citenamefont {Cerda},
  \citenamefont {Chaieb}, \citenamefont {Melo},\ and\ \citenamefont
  {Mahadevan}}]{cerda1999conical}%
  \BibitemOpen
  \bibfield  {author} {\bibinfo {author} {\bibfnamefont {E.}~\bibnamefont
  {Cerda}}, \bibinfo {author} {\bibfnamefont {S.}~\bibnamefont {Chaieb}},
  \bibinfo {author} {\bibfnamefont {F.}~\bibnamefont {Melo}}, \ and\ \bibinfo
  {author} {\bibfnamefont {L.}~\bibnamefont {Mahadevan}},\ }\href@noop {}
  {\bibfield  {journal} {\bibinfo  {journal} {Nature}\ }\textbf {\bibinfo
  {volume} {401}},\ \bibinfo {pages} {46} (\bibinfo {year} {1999})}\BibitemShut
  {NoStop}%
\bibitem [{\citenamefont {Walker}\ and\ \citenamefont
  {Seffen}(2018)}]{walker2018shape}%
  \BibitemOpen
  \bibfield  {author} {\bibinfo {author} {\bibfnamefont {M.~G.}\ \bibnamefont
  {Walker}}\ and\ \bibinfo {author} {\bibfnamefont {K.~A.}\ \bibnamefont
  {Seffen}},\ }\href@noop {} {\bibfield  {journal} {\bibinfo  {journal}
  {Thin-Walled Structures}\ }\textbf {\bibinfo {volume} {124}},\ \bibinfo
  {pages} {538} (\bibinfo {year} {2018})}\BibitemShut {NoStop}%
\bibitem [{\citenamefont {Cerda}\ \emph {et~al.}(2004)\citenamefont {Cerda},
  \citenamefont {Mahadevan},\ and\ \citenamefont {Pasini}}]{cerda2004elements}%
  \BibitemOpen
  \bibfield  {author} {\bibinfo {author} {\bibfnamefont {E.}~\bibnamefont
  {Cerda}}, \bibinfo {author} {\bibfnamefont {L.}~\bibnamefont {Mahadevan}}, \
  and\ \bibinfo {author} {\bibfnamefont {J.~M.}\ \bibnamefont {Pasini}},\
  }\href@noop {} {\bibfield  {journal} {\bibinfo  {journal} {Proceedings of the
  National Academy of Sciences}\ }\textbf {\bibinfo {volume} {101}},\ \bibinfo
  {pages} {1806} (\bibinfo {year} {2004})}\BibitemShut {NoStop}%
\bibitem [{\citenamefont {Cerda}\ and\ \citenamefont
  {Mahadevan}(2005)}]{cerda2005confined}%
  \BibitemOpen
  \bibfield  {author} {\bibinfo {author} {\bibfnamefont {E.}~\bibnamefont
  {Cerda}}\ and\ \bibinfo {author} {\bibfnamefont {L.}~\bibnamefont
  {Mahadevan}},\ }\href@noop {} {\bibfield  {journal} {\bibinfo  {journal}
  {Proceedings of the Royal Society of London A: Mathematical, Physical and
  Engineering Sciences}\ }\textbf {\bibinfo {volume} {461}},\ \bibinfo {pages}
  {671} (\bibinfo {year} {2005})}\BibitemShut {NoStop}%
\bibitem [{\citenamefont {Guven}\ and\ \citenamefont
  {M{\"u}ller}(2008)}]{guven2008paper}%
  \BibitemOpen
  \bibfield  {author} {\bibinfo {author} {\bibfnamefont {J.}~\bibnamefont
  {Guven}}\ and\ \bibinfo {author} {\bibfnamefont {M.~M.}\ \bibnamefont
  {M{\"u}ller}},\ }\href@noop {} {\bibfield  {journal} {\bibinfo  {journal}
  {Journal of Physics A: Mathematical and Theoretical}\ }\textbf {\bibinfo
  {volume} {41}},\ \bibinfo {pages} {055203} (\bibinfo {year}
  {2008})}\BibitemShut {NoStop}%
\bibitem [{\citenamefont {Jules}\ \emph {et~al.}(2019)\citenamefont {Jules},
  \citenamefont {Lechenault},\ and\ \citenamefont
  {Adda-Bedia}}]{jules2019local}%
  \BibitemOpen
  \bibfield  {author} {\bibinfo {author} {\bibfnamefont {T.}~\bibnamefont
  {Jules}}, \bibinfo {author} {\bibfnamefont {F.}~\bibnamefont {Lechenault}}, \
  and\ \bibinfo {author} {\bibfnamefont {M.}~\bibnamefont {Adda-Bedia}},\
  }\href@noop {} {\bibfield  {journal} {\bibinfo  {journal} {Soft Matter}\
  }\textbf {\bibinfo {volume} {15}},\ \bibinfo {pages} {1619} (\bibinfo {year}
  {2019})}\BibitemShut {NoStop}%
\bibitem [{\citenamefont {Abramowitz}\ and\ \citenamefont
  {Stegun}(1965)}]{abramowitz1965handbook}%
  \BibitemOpen
  \bibfield  {author} {\bibinfo {author} {\bibfnamefont {M.}~\bibnamefont
  {Abramowitz}}\ and\ \bibinfo {author} {\bibfnamefont {I.~A.}\ \bibnamefont
  {Stegun}},\ }\href@noop {} {\emph {\bibinfo {title} {Handbook of mathematical
  functions: with formulas, graphs, and mathematical tables}}},\ Vol.~\bibinfo
  {volume} {55}\ (\bibinfo  {publisher} {Courier Corporation},\ \bibinfo {year}
  {1965})\BibitemShut {NoStop}%
\bibitem [{\citenamefont {Silverberg}\ \emph {et~al.}(2015)\citenamefont
  {Silverberg}, \citenamefont {Na}, \citenamefont {Evans}, \citenamefont {Liu},
  \citenamefont {Hull}, \citenamefont {Santangelo}, \citenamefont {Lang},
  \citenamefont {Hayward},\ and\ \citenamefont
  {Cohen}}]{silverberg2015origami}%
  \BibitemOpen
  \bibfield  {author} {\bibinfo {author} {\bibfnamefont {J.~L.}\ \bibnamefont
  {Silverberg}}, \bibinfo {author} {\bibfnamefont {J.-H.}\ \bibnamefont {Na}},
  \bibinfo {author} {\bibfnamefont {A.~A.}\ \bibnamefont {Evans}}, \bibinfo
  {author} {\bibfnamefont {B.}~\bibnamefont {Liu}}, \bibinfo {author}
  {\bibfnamefont {T.~C.}\ \bibnamefont {Hull}}, \bibinfo {author}
  {\bibfnamefont {C.~D.}\ \bibnamefont {Santangelo}}, \bibinfo {author}
  {\bibfnamefont {R.~J.}\ \bibnamefont {Lang}}, \bibinfo {author}
  {\bibfnamefont {R.~C.}\ \bibnamefont {Hayward}}, \ and\ \bibinfo {author}
  {\bibfnamefont {I.}~\bibnamefont {Cohen}},\ }\href@noop {} {\bibfield
  {journal} {\bibinfo  {journal} {Nature Materials}\ }\textbf {\bibinfo
  {volume} {14}},\ \bibinfo {pages} {389} (\bibinfo {year} {2015})}\BibitemShut
  {NoStop}%
\bibitem [{\citenamefont {Benusiglio}\ \emph {et~al.}(2012)\citenamefont
  {Benusiglio}, \citenamefont {Mansard}, \citenamefont {Biance},\ and\
  \citenamefont {Bocquet}}]{benusiglio2012anatomy}%
  \BibitemOpen
  \bibfield  {author} {\bibinfo {author} {\bibfnamefont {A.}~\bibnamefont
  {Benusiglio}}, \bibinfo {author} {\bibfnamefont {V.}~\bibnamefont {Mansard}},
  \bibinfo {author} {\bibfnamefont {A.-L.}\ \bibnamefont {Biance}}, \ and\
  \bibinfo {author} {\bibfnamefont {L.}~\bibnamefont {Bocquet}},\ }\href@noop
  {} {\bibfield  {journal} {\bibinfo  {journal} {Soft Matter}\ }\textbf
  {\bibinfo {volume} {8}},\ \bibinfo {pages} {3342} (\bibinfo {year}
  {2012})}\BibitemShut {NoStop}%
\bibitem [{Sup()}]{Supp}%
  \BibitemOpen
  \href@noop {} {}\bibinfo {note} {See Supplemental Material at [URL will be
  inserted by publisher] for movies showing the indentation tests of soft
  (Movie S1) and stiff (Movie S2) all-mountain $f$-cones made of 4 mountain
  creases.}\BibitemShut {Stop}%
\bibitem [{\citenamefont {Klein}\ \emph {et~al.}(2007)\citenamefont {Klein},
  \citenamefont {Efrati},\ and\ \citenamefont {Sharon}}]{klein2007shaping}%
  \BibitemOpen
  \bibfield  {author} {\bibinfo {author} {\bibfnamefont {Y.}~\bibnamefont
  {Klein}}, \bibinfo {author} {\bibfnamefont {E.}~\bibnamefont {Efrati}}, \
  and\ \bibinfo {author} {\bibfnamefont {E.}~\bibnamefont {Sharon}},\
  }\href@noop {} {\bibfield  {journal} {\bibinfo  {journal} {Science}\ }\textbf
  {\bibinfo {volume} {315}},\ \bibinfo {pages} {1116} (\bibinfo {year}
  {2007})}\BibitemShut {NoStop}%
\bibitem [{\citenamefont {Kim}\ \emph {et~al.}(2012)\citenamefont {Kim},
  \citenamefont {Hanna}, \citenamefont {Byun}, \citenamefont {Santangelo},\
  and\ \citenamefont {Hayward}}]{kim2012designing}%
  \BibitemOpen
  \bibfield  {author} {\bibinfo {author} {\bibfnamefont {J.}~\bibnamefont
  {Kim}}, \bibinfo {author} {\bibfnamefont {J.~A.}\ \bibnamefont {Hanna}},
  \bibinfo {author} {\bibfnamefont {M.}~\bibnamefont {Byun}}, \bibinfo {author}
  {\bibfnamefont {C.~D.}\ \bibnamefont {Santangelo}}, \ and\ \bibinfo {author}
  {\bibfnamefont {R.~C.}\ \bibnamefont {Hayward}},\ }\href@noop {} {\bibfield
  {journal} {\bibinfo  {journal} {Science}\ }\textbf {\bibinfo {volume}
  {335}},\ \bibinfo {pages} {1201} (\bibinfo {year} {2012})}\BibitemShut
  {NoStop}%
\bibitem [{\citenamefont {Wu}\ \emph {et~al.}(2013)\citenamefont {Wu},
  \citenamefont {Moshe}, \citenamefont {Greener}, \citenamefont
  {Therien-Aubin}, \citenamefont {Nie}, \citenamefont {Sharon},\ and\
  \citenamefont {Kumacheva}}]{wu2013three}%
  \BibitemOpen
  \bibfield  {author} {\bibinfo {author} {\bibfnamefont {Z.~L.}\ \bibnamefont
  {Wu}}, \bibinfo {author} {\bibfnamefont {M.}~\bibnamefont {Moshe}}, \bibinfo
  {author} {\bibfnamefont {J.}~\bibnamefont {Greener}}, \bibinfo {author}
  {\bibfnamefont {H.}~\bibnamefont {Therien-Aubin}}, \bibinfo {author}
  {\bibfnamefont {Z.}~\bibnamefont {Nie}}, \bibinfo {author} {\bibfnamefont
  {E.}~\bibnamefont {Sharon}}, \ and\ \bibinfo {author} {\bibfnamefont
  {E.}~\bibnamefont {Kumacheva}},\ }\href@noop {} {\bibfield  {journal}
  {\bibinfo  {journal} {Nature Communications}\ }\textbf {\bibinfo {volume}
  {4}},\ \bibinfo {pages} {1586} (\bibinfo {year} {2013})}\BibitemShut
  {NoStop}%
\end{thebibliography}%

\end{document}